\documentclass[lettersize,journal]{IEEEtran}

\usepackage{xr}
\usepackage{graphics}
\usepackage{cite}
\usepackage{wrapfig}
\usepackage{multirow}
\usepackage{pdflscape}
\usepackage{colortbl}
\usepackage{array}
\usepackage{tikz} 
\usetikzlibrary{positioning}
\usepackage{xcolor} 
\usepackage{amsmath, amssymb, amsfonts}
\usepackage{subcaption}
\usepackage{graphicx}
\usepackage{float}
\usepackage{booktabs}
\usepackage{mathrsfs}
\usepackage[ruled,vlined]{algorithm2e}
\usepackage{orcidlink}
\usepackage[normalem]{ulem}
\useunder{\uline}{\ul}{}
\def\BibTeX{{\rm B\kern-.05em{\sc i\kern-.025em b}\kern-.08em
    T\kern-.1667em\lower.7ex\hbox{E}\kern-.125emX}}
\usepackage{pifont}
\newcommand{\cmark}{\ding{51}}%
\newcommand{\xmark}{\ding{55}}%

\usepackage{wasysym}

\newcommand{\blackCircle}{\CIRCLE}      
\newcommand{\halfBlackCircle}{\LEFTcircle}  
\newcommand{\whiteCircle}{\Circle}      

\begin{document}

\title{Countering Backdoor Attacks in Image Recognition: A Survey and Evaluation of Mitigation Strategies}

\author{Kealan Dunnett\orcidlink{0000-0002-0010-1499}, Reza Arablouei\orcidlink{0000-0002-6932-2900}, Dimity Miller\orcidlink{0000-0001-6312-8325}, Volkan Dedeoglu\orcidlink{0000-0002-2567-2423}, Raja Jurdak\orcidlink{0000-0001-7517-0782}
\thanks{\textit{(Corresponding author: Kealan Dunnett.)}}
\thanks{Kealan Dunnett, Dimity Miller, and Raja Jurdak are with the Queensland University of Technology, Brisbane, QLD 4000, Australia (e-mail: kealan.dunnett@hdr.qut.edu.au, \{d24.miller, r.jurdak\}@qut.edu.au).}
\thanks{D.M. acknowledges ongoing support from the QUT Centre for Robotics.}
\thanks{Reza Arablouei and Volkan Dedeoglu are with the Commonwealth Scientific and Industrial Research Organisation, Pullenvale, QLD 4069, Australia (e-mail: \{reza.arablouei, volkan.dedeoglu\}@csiro.au).}}


\maketitle

\begin{abstract}

The widespread adoption of deep learning across various industries has introduced substantial challenges, particularly in terms of model explainability and security. The inherent complexity of deep learning models, while contributing to their effectiveness, also renders them susceptible to adversarial attacks. Among these, backdoor attacks are especially concerning, as they involve surreptitiously embedding specific triggers within training data, causing the model to exhibit aberrant behavior when presented with input containing the triggers. Such attacks often exploit vulnerabilities in outsourced processes, compromising model integrity without affecting performance on clean (trigger-free) input data. In this paper, we present a comprehensive review of existing mitigation strategies designed to counter backdoor attacks in image recognition. We provide an in-depth analysis of the theoretical foundations, practical efficacy, and limitations of these approaches. In addition, we conduct an extensive benchmarking of sixteen state-of-the-art approaches against eight distinct backdoor attacks, utilizing three datasets, four model architectures, and three poisoning ratios. Our results, derived from 122,236 individual experiments, indicate that while many approaches provide some level of protection, their performance can vary considerably. Furthermore, when compared to two seminal approaches, most newer approaches do not demonstrate substantial improvements in overall performance or consistency across diverse settings. Drawing from these findings, we propose potential directions for developing more effective and generalizable defensive mechanisms in the future.

\end{abstract}

\begin{IEEEkeywords}
backdoor attack mitigation, cybersecurity, image recognition.
\end{IEEEkeywords}

\section{Introduction}\label{introduction}

In recent years, deep learning has seen remarkable advancements, driving its widespread adoption across diverse industries and academic fields. This rapid integration is evident in sectors such as healthcare, education, automotive, and logistics, where deep learning is increasingly utilized to foster innovation~\cite{pouyanfar2018survey}. A key factor in the success of deep learning is its ability to extract complex patterns from data. While this capability offers significant advantages, it also introduces substantial challenges related to explainability. Despite their strong predictive performance, deep learning models often lack transparency and interpretability, making it difficult to provide clear reasoning for specific predictions. 

The black-box nature of deep learning models has been shown to expose them to considerable security vulnerabilities~\cite{wang2019security}. In particular, classification models, such as those used in image recognition, have been demonstrated to be vulnerable to manipulation, with multiple instances of adversarial attacks successfully compromising their decision-making processes. For instance, the seminal work of~\cite{szegedy2013intriguing} highlights the vulnerability of image classification models to adversarial examples, showing that imperceptible perturbations applied to input images can induce significant misclassifications. The iconic ``Panda-Gibbon'' image, created using the method proposed in~\cite{szegedy2013intriguing} for generating adversarial perturbations, is a striking illustration of the fragility inherent in deep learning models, despite their sophistication. Since then, several other adversarial threats have been identified, affecting a wide range of learning tasks~\cite{wang2019security}.

\begin{figure}[t!]
    \centering
    \includegraphics[width=0.9\linewidth]{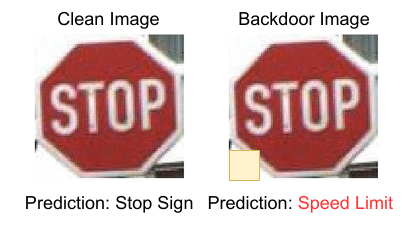}
    \caption{Example of a backdoor image (right) and its corresponding clean image (left). A yellow square, serving as the trigger, has been added to the bottom left corner of the backdoor image.}
    \label{fig:backdoor-example}
\end{figure}

In real-world scenarios, backdoor attacks represent a significant threat, especially in scenarios where classification outputs drive automated decision-making processes~\cite{gu2017badnets}. Backdoor attacks intentionally establish a relationship between a spurious feature within the input space, known as a trigger, and a specific classification outcome. Once this association is established, a compromised model will classify clean images (i.e., unaltered images) and backdoor images (i.e., images containing the trigger) differently. As a result, backdoor attacks undermine the integrity of a model's decision-making by inserting an unwanted behaviour, referred to as a backdoor task, without compromising its ability to correctly perform the original classification task of recognizing clean images~\cite{li2022backdoor}. In Figure~\ref{fig:backdoor-example}, we present both a clean image and a backdoor version, along with the classification results from a successfully compromised (backdoored) model. This backdoor attack can be practically executed by simply placing a yellow sticker on a stop sign. In an autonomous driving context, such an attack can have severe safety implications. 

\begin{figure*}[t]
    \centering
    \begin{subfigure}[b]{0.1\textwidth}
        \centering
        \includegraphics[width=\textwidth]{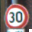}
        \caption{BadNets}
        \label{fig:backdoor-example-badnet}
    \end{subfigure}
    \begin{subfigure}[b]{0.1\textwidth}
        \centering
        \includegraphics[width=\textwidth]{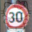}
        \caption{Blended}
        \label{fig:backdoor-example-blended}
    \end{subfigure}
    \begin{subfigure}[b]{0.1\textwidth}
        \centering
        \includegraphics[width=\textwidth]{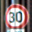}
        \caption{Signal}
        \label{fig:backdoor-example-signal}
    \end{subfigure}
    \begin{subfigure}[b]{0.1\textwidth}
        \centering
        \includegraphics[width=\textwidth]{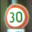}
        \caption{LF}
        \label{fig:backdoor-example-lf}
    \end{subfigure}
    \begin{subfigure}[b]{0.1\textwidth}
        \centering
        \includegraphics[width=\textwidth]{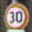}
        \caption{SSBA}
        \label{fig:backdoor-example-ssba}
    \end{subfigure}
    \begin{subfigure}[b]{0.1\textwidth}
        \centering
        \includegraphics[width=\textwidth]{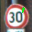}
        \caption{IAB}
        \label{fig:backdoor-example-IAB}
    \end{subfigure}
    \begin{subfigure}[b]{0.1\textwidth}
        \centering
        \includegraphics[width=\textwidth]{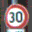}
        \caption{BPP}
        \label{fig:backdoor-example-bpp}
    \end{subfigure}
    \begin{subfigure}[b]{0.1\textwidth}
        \centering
        \includegraphics[width=\textwidth]{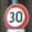}
        \caption{WaNet}
        \label{fig:backdoor-example-WaNet}
    \end{subfigure}
    \begin{subfigure}[b]{0.1\textwidth}
        \centering
        \includegraphics[width=\textwidth]{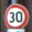}
        \caption{LIRA}
        \label{fig:backdoor-example-lira}
    \end{subfigure}
\caption{Examples of different backdoor triggers used in the literature. Note that while IAB adds a local patch to each image, its position and scale can vary across images.}
\label{fig:backdoor-example2}
\end{figure*}


To successfully execute a backdoor attack on a model, the adversary must compromise the training pipeline of the target model. This makes backdoor attacks particularly dangerous when model training is outsourced to third parties, such as through cloud-based Machine-Learning-as-a-Service platforms~\cite{liu2019abs}. In such cases, an adversary can manipulate the training data or procedure, injecting backdoor tasks without the victim's awareness. The widespread use of pre-trained model weights in the deep learning community exacerbates this threat. For instance, a recent industry survey~\cite{grosse2023towards} has revealed that 48.1\% of participants utilize third-party weights for model training. Moreover, popular machine learning platforms such as Hugging Face\footnote{https://huggingface.co/} allow users to download thousands of off-the-shelf pre-trained models, further amplifying the potential risk of backdoor attacks.

To address the challenges posed by the hard-to-explain nature of current deep learning approaches, machine learning security has emerged as a critical area of research~\cite{yuan2019adversarial}. Although it does not directly solve the problem of explainability, machine learning security seeks to develop new methods to protect against known threats, ensuring that deep learning can be safely deployed. Within this field, an emerging body of work focuses on developing defensive strategies specifically designed to counter backdoor attacks. These strategies aim to mitigate risks by removing the backdoor task from a model while preserving its ability to classify clean inputs. However, despite significant advancements in this field, the consistency and reliability of current proposals across diverse settings remain uncertain due to practical limitations. Many proposed approaches are evaluated using a limited scope of attacks, datasets, model architectures, and data availability conditions. This raises questions about their generalizability and effectiveness in diverse real-world scenarios. 

In this work, we critically analyse existing research on backdoor mitigation, distinguishing it from other defensive approaches such as identifying poisoned training examples or synthesizing backdoored inputs. Our analysis is centered on works designed for image classification, although backdoor attacks are also relevant in other applications such as natural language processing~\cite{sheng2022survey} and other computer vision tasks such as object detection~\cite{chan2022baddet} and semantic segmentation~\cite{li2021hidden}. We conduct a comprehensive survey and critical analysis of existing mitigation approaches. Unlike other surveys on this topic, we offer a detailed summary of the prevalent approaches used to address backdoor attacks, along with their main assumptions and limitations. Furthermore, we experimentally evaluate the majority of the discussed works against a diverse range of backdoor attacks, covering eight distinct types. Our evaluations, comprising 122,236 individual experiments, spanning three different datasets, four model architectures, and three distinct data availability settings. Our benchmarking results reveal several key findings that offer valuable insights to inform future research directions.


\textit{Existing Surveys and Benchmarks}: There exist several surveys on backdoor attack mitigation. We provide a comparative summary of these works in Table~\ref{tab:survey-compare}. Most of these surveys take a broader approach than ours, which often results in a less detailed analysis of the methods currently employed within the image classification domain. For instance, \cite{sheng2022survey}, \cite{cheng2023backdoor}, and \cite{zhao2024survey} concentrate on language tasks, while~\cite{le2024comprehensive} and \cite{yan2023backdoor} focus on face and voice recognition tasks, respectively. Additionally, \cite{wan2024data} specifically surveys mitigation strategies devised for federated learning.

\begin{table}[t]
    \centering
    \caption{Comparison of our work with existing related surveys. Note: \textit{\# Evaluated} refers to the number of mitigation proposals benchmarked.}
    \scalebox{0.83}{
        \begin{tabular}{cccccc}
        \toprule
             \multirow{2}{*}{Reference} & Year & \multirow{2}{*}{Domain} & \multirow{2}{*}{Defensive Tasks} & Experimental & \multirow{2}{*}{\# Evaluated} \\
             & & & & Evaluation & \\ 
        \toprule
        \cite{sheng2022survey} & 2022 & Text & Various & \xmark & N/A \\
        \cite{cheng2023backdoor} & 2023 & Text & Various & \xmark & N/A \\ 
        \cite{zhao2024survey} & 2024 & Text & Various & \xmark & N/A \\ 
        \cite{yan2023backdoor} & 2023 & Voice & Various & \xmark & N/A \\
        \cite{wan2024data} & 2024 &  \begin{tabular}{c} Federated \\ Learning \end{tabular} & Various & \xmark & N/A \\
        \cite{le2024comprehensive} & 2024 & Image & Various & \xmark & N/A \\
        \cite{li2022backdoor} & 2022 & Image & Various & \xmark & N/A \\
        \cite{wu2022backdoorbench} & 2022 & Image & Various & \cmark & 4 \\
        \textbf{Ours} & 2024 & Image & Mitigation & \cmark & 16 \\
        \hline
        \end{tabular}
    }
    \label{tab:survey-compare}
\end{table}

The two works most similar to ours are \cite{li2022backdoor} and \cite{wu2022backdoorbench}. In \cite{li2022backdoor}, multiple classification tasks are surveyed, but the analysis of methods specific to image classification is not detailed. While~\cite{wu2022backdoorbench} focuses on image classification, its primary aim is the development of a benchmarking tool. Although~\cite{wu2022backdoorbench} offers an extensive evaluation of various defensive approaches, only four of the nine evaluated methods perform mitigation, with the others employing different defensive strategies. Moreover, the four mitigation approaches assessed by~\cite{wu2022backdoorbench} are not considered state-of-the-art, particularly in light of more recent results presented in \cite{zhu2024npd} and \cite{zeng2022i-bau}.

\section{Background and Preliminaries}

In this section, we introduce key foundational concepts relevant to our survey and establish a consistent set of notations that will be utilized throughout the paper. 

\subsection{Notation} \label{notation}

Here, we provide a concise overview of the general notation that is used in the subsequent sections. In Table~\ref{tab:notation}, we list the common symbols that are referenced throughout the paper.

\begin{table}[t]
    \centering
    \caption{The list of common symbols.}
    \scalebox{0.95}{
    \begin{tabular}{cl}
        \toprule
        Symbol & Meaning \\
        \toprule
        $\theta$ & Model Parameters  \\
        $\varphi$ & Parameters in $\theta$ associated with feature extractor \\
        $\omega$ & Parameters in $\theta$ associated with the linear classifier \\
        $\phi$ & Filter matrix of a convolutional layer \\
        $\xi$ & Parameter perturbation applied to $\theta$ \\
        $\delta$ & Input perturbation applied to $x$ \\
        $\epsilon$ & Perturbation budget \\
        $\lambda$ & Loss Hyperparameter \\
        $p$ & Norm type used to define $\|\cdot\|_{p}$ \\
        \hline
        $\mathcal{X} \subseteq \mathbb{R}^{h \times w \times c}$ & Input space with $c$ channels, $w$ width and $h$ height \\
        $x \in \mathcal{X}$ & Clean input image \\ 
        $\hat{x} \in \mathcal{X}$ & Backdoor image \\
        $\rho$ & Trigger Pattern applied to $x$ to produce $\hat{x}$ \\
        $\mathcal{M} = \{0, 1, \dots , m\}$ & Label space \\
        $y \in \{0, 1\}^{m}$ & Correct label associated with $x$ (One-hot encoded)\\
        $\hat{y} \in \{0, 1\}^{m}$ & Target label associated with $\hat{x}$ (One-hot encoded)\\
        $(x, y) \in \mathcal{D}_{t}$ & Training Dataset \\
        $(x, y) \in \mathcal{D}_{c}$ & Clean Dataset \\
        $(\hat{x}, \hat{y}) \in \mathcal{D}_{b}$ & Backdoor Dataset \\
        $(x, y) \in \mathcal{D}_{m}$ & Mitigation Dataset \\
        $(x, y) \in \mathcal{D}_{v}$ & Validation Dataset \\
        \hline
        $z \in \mathbb{Q}^{m}$ & Logit output of a model \\
        $a \in [0,1]^{m}$ & Softmax output of a model \\
        \hline
    \end{tabular}}
    \label{tab:notation}
\end{table}

\subsubsection{Loss Functions} Fundamental to any training process is the use of a loss function. Within image classification, \textit{cross-entropy} loss, denoted as $\mathcal{L}_{\mathrm{CE}}$, is most commonly used. Using the predicted softmax probability vector $a$ for a given input $x$ and its corresponding one-hot true label vector $y$, $\mathcal{L}_{\mathrm{CE}}$ associated with this data instance is expressed as

\begin{equation} \label{ce-loss:single}
    \mathcal{L}_{\mathrm{CE}}(x, y, \theta) = -y^\intercal \log(a).
\end{equation}

In addition, the cross-entropy loss for a dataset $(x, y) \in \mathcal{D}$ is calculated as

\begin{equation}
\begin{aligned}
    \mathcal{L}_{\mathrm{CE}}(\mathcal{D}, \theta) =- &\frac{1}{|\mathcal{D}|} \sum_{(x, y) \in \mathcal{D}} y^\intercal \log(a).
\end{aligned}
\end{equation}

\subsubsection{Neural Network} A neural network is a parameterized function $f(x,\theta)$ that maps an input $x$ to an output $y$, where $\theta$ represents the network's parameters. In an $m$-class classification task, the output consists of softmax probabilities for each class. Given a training dataset $\mathcal{D}_{t}$, $\theta$ is optimized to minimize an aggregate loss function, typically $\mathcal{L}_{\mathrm{CE}}$ for image classification, as

\begin{equation} \label{general-training-objective}
    \min_{\theta} \mathcal{L}_{\mathrm{CE}}(\mathcal{D}_{t}, \theta).
\end{equation}
Stochastic gradient descent or its variants is commonly used for optimization. 

Neural network architectures vary widely, with convolutional neural networks (CNNs) being prevalent for image classification. CNNs can often be decomposed into a feature extractor and a linear classifier. Letting $\varphi$ and $\omega$ represent the parameters of these components, respectively, we have $\omega \cup \varphi \subseteq \theta$ and $\omega \cap \varphi = \emptyset$.

\subsection{Backdoor Attacks} \label{prelim:backdoor-attacks}

To execute a backdoor attack, various methods have been proposed in the literature, typically involving the creation of two datasets $\mathcal{D}_{c}$ (clean) and $\mathcal{D}_{b}$ (backdoor), which combine to form $\widetilde{\mathcal{D}}_{t}$, a poisoned variant of the original training dataset $\mathcal{D}_{t}$. A critical consideration in this process is the poisoning ratio, which represents the proportion of backdoor to clean data (i.e., $\frac{\|\mathcal{D}_{b}\|}{\|\mathcal{D}_{c}\|}$), balancing the trade-off between the attack's stealth and effectiveness. The clean dataset $\mathcal{D}_{c}$ consists of the original unaltered inputs $x$ and their associated labels $y$. To generate the backdoor dataset $\mathcal{D}_{b}$, a backdoor function $B(x, \rho) \rightarrow \hat{x}, \hat{y}$ is used, where $\rho$ denotes the trigger pattern added to $x$ to produce the backdoored input $\hat{x}$. In  targeted backdoor attacks, which are the most extensively studied in the literature, the label $\hat{y}$ is typically assigned a predefined value. Although alternative adversarial objectives, such as the all-target objective described in \cite{zhao2020bridging}, have been explored, targeted backdoor attacks remain the primary focus within this domain.
Using $\widetilde{\mathcal{D}}_{t}$, $\theta$ is optimized to accurately classify both $\mathcal{D}_{c}$ and $\mathcal{D}_{b}$.
Recent works, such as LIRA~\cite{doan2021lira}), have introduced specialized training procedures tailored to enhance the effectiveness of backdoor attacks, which, while relevant, fall outside the scope of this work.

\begin{figure*}[t]
    \centering
    \begin{subfigure}[b]{0.4\linewidth}
        \centering
        \includegraphics[width=\linewidth]{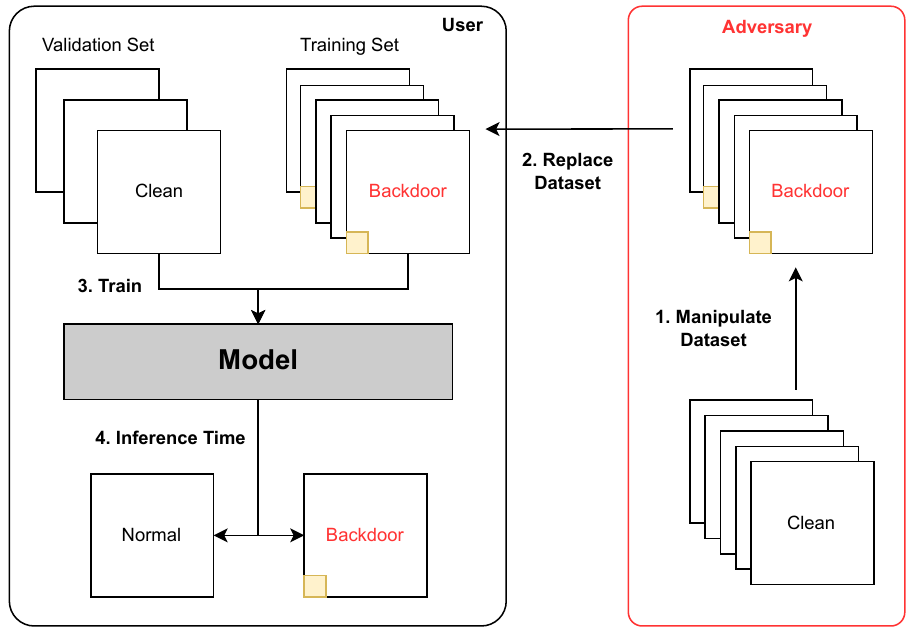}
        \caption{Data Poisoning (DP)}
        \label{fig:poisoning-threat}
    \end{subfigure}
    \begin{subfigure}[b]{0.48\linewidth}
        \centering
        \includegraphics[width=\linewidth]{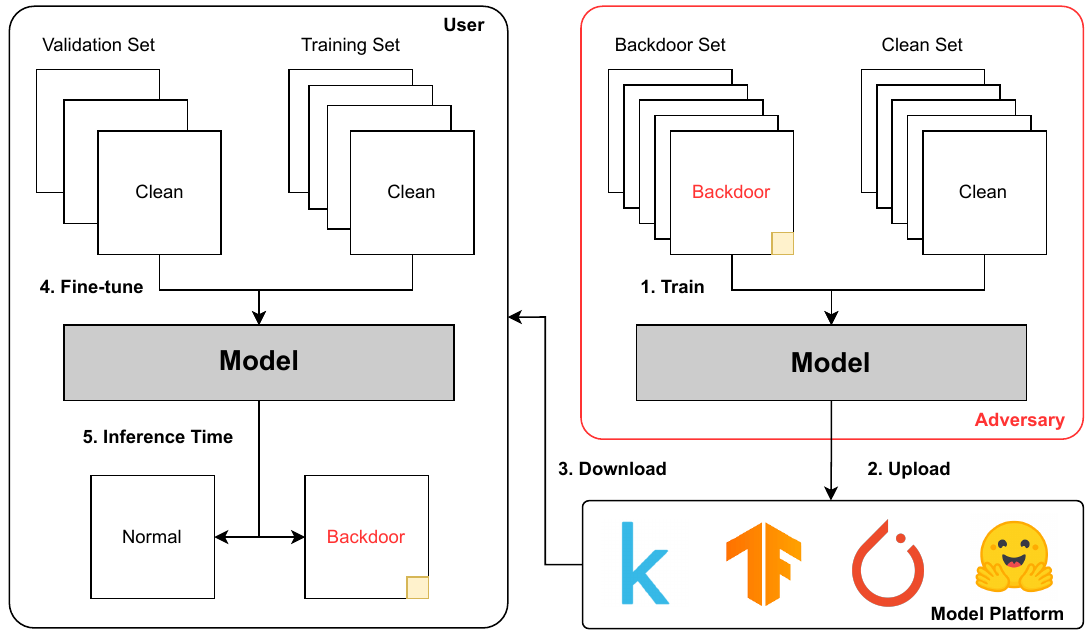}
        \caption{Outsourced Training (OT)}
        \label{fig:outsource-threat}
    \end{subfigure}
\caption{Threat models considered by existing backdoor attacks.}
\label{fig:threat-models}
\end{figure*}

Current backdoor attacks are primarily analyzed within two main threat models, as depicted in Figure~\ref{fig:threat-models}. The first, the \textit{Data Poisoning} threat model, assumes that the adversary's capabilities are limited to modifying the training data, allowing them to replace $\mathcal{D}_{t}$ with $\widetilde{\mathcal{D}}_{t}$. The second, more potent adversarial scenario is the \textit{Outsourced Training} threat model, where the adversary has full control over the entire training process. In this setting, the adversary can not only substitute $\mathcal{D}_{t}$ with $\widetilde{\mathcal{D}}_{t}$, but also employ an arbitrary non-auditable training procedure.
Using these threat models, along with additional characteristics introduced in studies such as \cite{wu2022backdoorbench}, we categorize prominent backdoor attacks relevant to image classification in Table~\ref{tab:attack-details}. While not exhaustive, this table highlights the most significant works in this area.  

\begin{table}[t]
    \centering
    \caption{Categorization of state-of-the-art backdoor attacks based on key characteristics and the training procedure employed by the adversary. DP: Data Poisoning, OT: Outsource Training.}
    \begin{tabular}{cccccc}
        \toprule
        Ref & Name & \multicolumn{3}{c}{Trigger Characteristics} & Threat\\
        & & Coverage & Consistency & Mode & Model \\
        \toprule
        \cite{gu2017badnets} & BadNets & Local & Static & Replacement  & DP \\
        \cite{chen2017targeted} & Blended & Global & Static & Additive  & DP \\
        \cite{barni2019new} & Signal & Global & Static & Additive  & DP \\
        \cite{zeng2021rethinking} & LF & Global & Static & Additive  & DP \\
        \cite{li2021invisible} & SSBA & Global & Dynamic & Additive  & DP \\
        \cite{nguyen2020input} & IAB & Local* & Dynamic & Replacement  & OT \\
        \cite{wang2022bppattack} & BPP & Global & Dynamic & Additive  & OT \\
        \cite{nguyen2021wanet} & WaNet & Global & Dynamic & Warping  & OT \\
        \cite{doan2021lira} & LIRA & Global & Dynamic & Additive & OT \\
        \hline
    \end{tabular}
\vspace{6pt}
\caption*{\footnotesize *IAB generates triggers that typically cause minor localized pixel changes, the trigger pattern can have a global impact on the image.}
    \label{tab:attack-details}
\end{table}

\subsubsection{Trigger Characteristics}

Trigger characteristics pertain to the type of trigger $\rho$ used by the adversary, encompassing coverage, consistency and modification mode, as defined in~\cite{wu2022backdoorbench}. Below, we briefly discuss these characteristics with reference to the works listed in Table~\ref{tab:attack-details}. 

\paragraph{Coverage} Trigger coverage refers to the extent of modification when $\rho$ is integrated into the $x$ by $B$~\cite{wu2022backdoorbench}. In Figure~\ref{fig:backdoor-example2}, we provide examples of backdoor images generated by the considered attacks. Coverage is typically classified as global or local. Global coverage implies that $\rho$ affects a significant portion of $x$. For instance, the Blended attack~\cite{chen2017targeted} alters $x$ by incorporating a trigger image (e.g., a picture of Hello Kitty) with a blending ratio that controls its transparency. In contrast, local triggers modify only a small portion of $x$. Among the attacks considered, BadNets~\cite{gu2017badnets} and IAB~\cite{nguyen2020input} utilize local triggers. In the case of BadNets, a small $n \times n$ pixel pattern (e.g., a $3 \times 3$ white square) is inserted into images at a fixed position (e.g., at the bottom left-hand corner).

\paragraph{Consistency} Trigger consistency refers to whether the same $\rho$ is used across $\mathcal{D}_{b}$. When $\rho$ is fixed, the attack is deemed static. For example, the BadNets attack inserts the same $n \times n$ pixel pattern into the same position in each image. However, the success of works such as \cite{wang2019neural} in identifying static triggers used by BadNets has led to the preference for dynamic triggers in recent works. More specifically, dynamic methods make $\rho$ dependent on $x$. For example, IAB generates a unique $\rho$ for each $(x, y) \in \mathcal{D}_{c}$. Unlike other dynamic methods such as SSBA, BPP, and LIRA, which synthesize barely perceptible patterns, IAB generates patterns similar to those used in BadNets. Moreover, the triggers generated by IAB for different images are designed to be non-reusable, meaning the trigger for $x$ does not work for $x'$.


\begin{figure}[t!]
    \centering
    \begin{subfigure}[b]{0.1\textwidth}
        \centering
        \includegraphics[width=\textwidth]{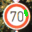}
    \end{subfigure}
    \begin{subfigure}[b]{0.1\textwidth}
        \centering
        \includegraphics[width=\textwidth]{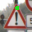}
    \end{subfigure}
    \begin{subfigure}[b]{0.1\textwidth}
        \centering
        \includegraphics[width=\textwidth]{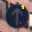}
    \end{subfigure}
    \begin{subfigure}[b]{0.1\textwidth}
        \centering
        \includegraphics[width=\textwidth]{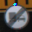}
    \end{subfigure}
\caption{Examples of different IAB trigger patterns.}
\label{fig:iab-trigger-examples}
\end{figure}

\paragraph{Mode} Trigger modification mode refers to how $\rho$ is applied to $x$. The two most common modes are additive and replacement. Attacks using the replacement mode substitute parts of $x$ with $\rho$. For example, BadNets replaces the pixels in the $n \times n$ region of $x$ that overlaps with $\rho$. in contrast, additive attacks add $\rho$ to $x$ (i.e., $\hat{x} = x + \rho$). The exception to this categorization is the WaNet attack, where the elastic image warping method utilized in this attack cannot be classified as either additive or replacement.

\subsection{Backdoor Mitigation}

The landscape of defense against backdoor attacks encompasses several sub-problems, addressed by different works. For instance, some works propose methods that identify backdoor samples within the poisoned dataset $\widetilde{\mathcal{D}}_{t}$ \cite{gao2019strip}. Another common task is backdoor synthesis, which involves generating the backdoor dataset $\mathcal{D}_{b}$ from the clean dataset $\mathcal{D}_{c}$ given the model parameters $\theta$~\cite{liu2019abs}. Such works attempt to model the trigger distribution used by the adversary.

In this work, we concentrate on backdoor mitigation. Distinct from other defensive strategies, backdoor mitigation seeks to remove backdoor behavior from the model while preserving its original classification capability~\cite{liu2018finepruning}. Although several works address backdoor mitigation, inconsistencies often arise concerning the threat model considered by each work. To ensure a fair comparison, we focus on proposals that adopt a set of assumptions consistent with the strongest adversarial scenario, specifically, the Outsourced Training threat model outlined in section~\ref{prelim:backdoor-attacks}. In this setting, the defender, who performs the mitigation, has access to $\theta$ and a small set of clean mitigation data $\mathcal{D}_{m}$. Crucially, the defender does not have access to any backdoor data $\mathcal{D}_{b}$. In addition, our survey only considers works published in prominent peer-reviewed venues, particularly, CORE\footnote{\url{https://www.core.edu.au/icore-portal}} ranked A/A* conferences and SJR\footnote{\url{https://www.scimagojr.com/journalrank.php}} ranked Q1 journals.             

In the the following sections, we present a comprehensive analysis of existing backdoor mitigation approaches within the image classification domain. We categorize these works based on their underlying methodologies, as summarized in Table~\ref{tab:mitigation_methods}.

\begin{table}[t]
\centering
\caption{Categorization of the surveyed backdoor mitigation approaches.}
\begin{tabular}{cccc}
\toprule
\textbf{Ref} & \textbf{Name} & \textbf{Category} & \textbf{Type} \\
\toprule
    \cite{liu2018finepruning} & FP & \multirow{8}{*}{Pruning} & \multirow{3}{*}{Metric} \\
    \cite{zheng2022bnp} & BNP & & \\
    \cite{zheng2022clp} & CLP & & \\
    \cline{4-4}
    \cite{wu2021anp} & ANP & & \multirow{3}{*}{Masking} \\
    \cite{chai2022awn} & AWN & & \\
    \cite{li2023rnp} & RNP & & \\
    \cline{4-4}
    \cite{wang2023mm} & MM-BD & & \multirow{2}{*}{Additive} \\
    \cite{zhu2024npd} & NPD & & \\
    
    \cline{4-4}
    \cite{wang2019nc} & NC* & & Synthesis Unlearn \\
    \cline{1-4}
    \cite{qiao2019mesa} & MESA & \multirow{9}{*}{Fine-Tuning} & \multirow{2}{*}{Synthesis Unlearn} \\
    \cite{liu2022baeraser} & BAERASER & & \\
    
    \cline{4-4}
    \cite{min2024fst} & FST & & \multirow{2}{*}{Traditional} \\
    \cite{zhu2023ft-sam} & FT-SAM & & \\
    \cline{4-4}
    \cite{li2021nad} & NAD & & \multirow{2}{*}{Knowledge Distillation} \\
    \cite{pang2023bcu} & BCU & & \\
    \cline{4-4}
    \cite{zeng2022i-bau} & i-BAU & & \multirow{3}{*}{Adversarial Training} \\
    \cite{mu2023pbe} & PBE & & \\
    \cite{wei2023sau} & SAU & & \\
    \hline
\end{tabular}
\vspace{6pt}
\caption*{\footnotesize *While NC is grouped with MESA and BAERASER, it is indeed a pruning approach.}
\label{tab:mitigation_methods}
\end{table}

\section{Model Pruning} \label{model-pruning}

\begin{figure*}[t]
    \centering
    \includegraphics[width=0.85\linewidth]{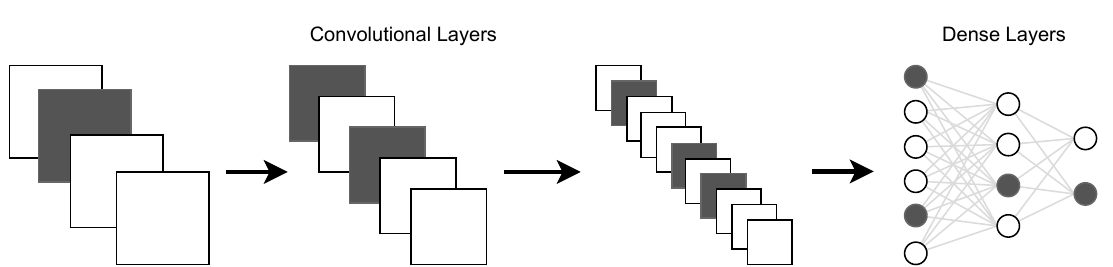}
    \caption{Visual representation of how a model can be segmented using the hypothesis introduced in \cite{gu2017badnets}. Grey and white distinguish between the backdoor and clean components respectively.}
    \label{fig:model-segmentation}
\end{figure*}

When backdoor attacks were first introduced for image classification, \cite{gu2017badnets} hypothesized that backdoored models extract two distinct \textit{feature} sets. They posited that certain model components, such as convolutional and dense layers, become associated either with the main task or the backdoor task. Consequently, when the model is presented with $\hat{x}$, the components linked to the backdoor task contribute to its classification as $\hat{y}$ rather than $y$. In Figure~\ref{fig:model-segmentation}, we visually represent how a model can be decomposed according to this hypothesis. 

Historically model pruning has been employed to identify and remove redundant model components, thereby improving inference efficiency~\cite{he2023structured}. Building upon the hypothesis that model components can be segmented as described, several studies have explored the application of model pruning for backdoor mitigation. Pruning-based approaches aim to identify and eliminate model components associated with backdoor behavior. To achieve this, various strategies have been adopted, including metric-based, masking-based, and additive techniques. In this section, we provide a comprehensive review of these subcategories, along with a comparative analysis of the approaches within them. 

\subsection{Metric-based Pruning}  

Works adopting a metric-based approach aim to directly quantify the contribution of each model component to the backdoor task. By applying a defined metric, these works distinguish between clean and backdoor components based on their respective metric values. 

\subsubsection{FP} The work of \cite{liu2018finepruning} is considered a pioneering effort in backdoor mitigation. In their initial investigation, \cite{liu2018finepruning} compare the channel-wise average activation produced by $\mathcal{D}_{c}$ and $\mathcal{D}_{b}$ in the final convolutional layer. Their analysis reveals that backdoor components, specifically a small subset of filters within this layer, are only activated by $\mathcal{D}_{b}$ (see \cite[Fig. 4]{liu2018finepruning}). 

Building on this observation, \cite{liu2018finepruning} propose pruning the final convolutional layer based on the average channel-wise activation given $\mathcal{D}_{m}$, the mitigation dataset available to the defender. The approach involves iteratively pruning filters with the lowest average activation until the accuracy on a validation dataset $\mathcal{D}_{v}$, which is segmented from $\mathcal{D}_{m}$ prior to pruning, falls below a defined threshold. After pruning, \cite{liu2018finepruning} fine-tune $\theta$ using $\mathcal{D}_{m}$ to recover any performance lost.    

\subsubsection{BNP} Similar to FP~\cite{liu2018finepruning}, \cite{zheng2022bnp} examine the activation values produced by a backdoored model. However, \cite{zheng2022bnp} compare the pre-activation distribution of clean and backdoor components. Pre-activation refers to the activation values before any non-linear transformation, e.g., ReLU, is applied. Their analysis reveals that clean components typically follow a unimodal distribution, while backdoor components exhibit a bimodal distribution (see \cite[Fig. 2(a)-(b)]{zheng2022bnp}). 

As an extension of the above analysis, \cite{zheng2022bnp} compare the batch normalization (BN) statistics ($\mu_{bn}$ and $\sigma_{bn}$) tracked in the proceeding layer with the pre-activation statistics of $\mathcal{D}_{c}$ ($\mu_{c}$ and $\sigma_{c}$). Their comparison shows that $\mu_{bn}$ and $\sigma_{bn}$ are biased relative to $\mu_{c}$ and $\sigma_{c}$, a consequence of $\mu_{bn}$ and $\sigma_{bn}$ stemming from the bimodal input distribution produced by $\widetilde{\mathcal{D}}_{t}$ (see \cite[Fig. 2(c)-(d)]{zheng2022bnp}). 

To leverage this characteristic, \cite{zheng2022bnp} calculate the Kullback-Leibler (KL) divergence between $\mathcal{N}(\mu_{bn}, \sigma_{bn})$ and $\mathcal{N}(\mu_{m}, \sigma_{m})$. The KL divergence is calculated for each filter within each convolutional layer. The set of KL-divergence values for the $l^{\text{th}}$ layer, denoted as $K_{l} = \{k_{1}, k_{2} \dots ,k_{n}\}$, is then used to determine a layer-specific pruning threshold $\tau_{l}$, calculated as 
\begin{equation}
    \tau_{l} = \bar{K}_{l} + \lambda s_{l},
\end{equation}  
where $\bar{K}_{l}$ and $s_l$ are the mean and standard deviation of $K_{l}$, and $\lambda$ is a hyperparameter selected by the defender. The filters are then pruned base on $\tau_{l}$. However, it is important to note that BNP assumes that a subset of filters exhibits biased BN statistics compared to the pre-activation statistics of $\mathcal{D}_{c}$. While this characteristic is demonstrated for a single filter in~\cite[Fig. 2(c)-(d)]{zheng2022bnp}, the distribution of $K_{l}$ across all layers is not provided, leaving limited evidence to support the assertion that filters in each layer can be distinctly separated using the proposed metric.

\subsubsection{CLP} Unlike other metric-based methods, \cite{zheng2022clp} propose using the \textit{Lipschitz} constant associated with each filter matrix $\phi$ to guide pruning. For the $i^{\text{th}}$ filter in the $l^{\text{th}}$ layer, the upper bound channel Lipschitz constant (UCLC) of $\phi_{l,i}$ is estimated as the largest singular value from the spectral decomposition of $\phi_{l,i}$. Note that $\phi_{l,i}$ is reshaped such that~$\phi_{l,i}\in\mathbb{R}^{c\times(hw)}$, where $c$, $h$, and $w$ represent the channel, height, and width dimensions of $\phi_{l,i}$. 

Operating under the hypothesis that backdoor components exhibit distinct activation patterns for $\mathcal{D}_{c}$ and $\mathcal{D}_{b}$, \cite{zheng2022clp} argue that the UCLC can effectively quantify this difference without needing access to either dataset. To validate this idea, \cite{zheng2022clp} introduces the trigger activation change (TAC), which quantifies the average activation difference between $\mathcal{D}_{c}$ and $\mathcal{D}_{b}$ for the $i^{\text{th}}$ filter in the $l^{\text{th}}$ layer, as 

\begin{equation} \label{clp:tac}
    \frac{1}{|\mathcal{D}_{c}|} \sum_{(x, y) \in \mathcal{D}_{c}, (\hat{x}, \hat{y}) \in \mathcal{D}_{b}} \|f_{l,i}(x, \theta) - f_{l,i}(\hat{x}, \theta) \|_{2},
\end{equation}

where $(x, y) \in \mathcal{D}_{c}, (\hat{x}, \hat{y}) \in \mathcal{D}_{b}$ is a clean and backdoor image pair. By plotting UCLC and TAC against each other, \cite{zheng2022clp} demonstrate a strong positive correlation between the two metrics (see \cite[Fig. 3]{zheng2022clp}). They subsequently suggest that UCLC can reliably quantify the sensitivity of each filter to $\rho$, allowing for a distinction between clean and backdoor components. For model pruning, \cite{zheng2022clp} adopt the same layer-based thresholding approach as BNP~\cite{zheng2022bnp}.

In contrast to BNP, \cite{zheng2022clp} examine the relationship between TAC and UCLC across multiple layers (see \cite[Fig. 3]{zheng2022clp}). However, while these metrics are well-correlated, it is important to note that most filters display some sensitivity to $\rho$ (i.e., most filters have a non-zero TAC value). In particular, filters do not form two distinct clusters, indicating that pruning may not be the most effective strategy.

\subsection{Masking-based Pruning}

Beyond metric-based approaches, some pruning methods focus on learning a parameter mask $\mathbf{m}$ that, when applied to $\theta$, effectively removes the backdoor behaviour. These methods formulate an objective function and use optimization techniques to find an optimal mask $\mathbf{m}$. The mask is applied using the Hadamard (element-wise) product as $\mathbf{m} \odot \theta$. Hence, when an entry of $\mathbf{m}$ is zero, the corresponding entry in $\theta$ is pruned. Additionally, since $\mathbf{m}$ is used to mask convolutional filters, a mask value is learned for each filter in each layer.  


\subsubsection{ANP}

In \cite{wu2021anp}, the authors initially frame the backdoor mitigation task as a masking problem, where they examine the impact of a perturbation $\xi$ applied to $\theta$ on the classification error over $\mathcal{D}_{m}$ for both clean and backdoored models. The perturbation set $\xi$ consists of two subsets: $\xi_w$, applied to the weights ($w$) in $\theta$, and $\xi_b$, applied to the biases ($b$) in $\theta$. Therefore, $\xi = \xi_w \: \cup \: \xi_b$ and $\theta = w \: \cup \: b$. To determine $\xi$, the following optimization problem is solved

\begin{equation} \label{anp-max}
    \max_{\xi_w, \xi_b \in [-\epsilon, \epsilon]} \mathcal{L}_\mathrm{CE}(\mathcal{D}_{m}, \left[(1 + \xi_w) \odot w \: \cup \: (1 + \xi_b) \odot b \right] ),
\end{equation} 
where $\epsilon$ constrains the values of $\xi$. In \cite{wu2021anp}, it is shown that a backdoored model exhibits higher classification error (see \cite[Fig. 1(a)]{wu2021anp}). It is also observed that classification errors made by the backdoored model are biased towards the target class (see \cite[Fig. 1(b)]{wu2021anp}). Based on these results, \cite{wu2021anp} hypothesise that $\xi$ targets backdoor components of the model. To leverage this characteristic, they propose solving the following minimax optimization problem

\begin{equation} \label{anp-min}
\begin{split}
    &\min_{\mathbf{m} \in [0, 1]}\biggl\{ \lambda \mathcal{L}_\mathrm{CE}\left(\mathcal{D}_{m}, \left[ (\mathbf{m} + \xi_w) \odot w, b \right] \right)\\
    &+ \max_{\substack{\xi_w, \xi_b \\ \in [-\epsilon, \epsilon]}} (1 - \lambda) \mathcal{L}_\mathrm{CE}\left(\mathcal{D}_{m}, \left[ (\mathbf{m} + \xi_w) \odot w, (1 + \xi_b) \odot b \right] \right)\biggr\}
\end{split}
\end{equation}
where $\lambda \in [0, 1]$ is a trade-off coefficient chosen by the defender. Solving~\eqref{anp-min} yields a perturbation $\xi$ that maximises the classification error of $\mathcal{D}_{m}$. At the same time, the outer minimisation results in a mask $\mathbf{m}$ that minimises the classification error of $\mathcal{D}_{m}$ given $\xi$. In \cite{wu2021anp}, they alternate between solving the inner and outer sub-problems multiple times. Once a solution for $\mathbf{m}$ is found, it is binarised using a threshold value or a fixed pruning fraction. 

\subsubsection{AWM}

The robustness of ANP~\cite{wu2021anp} in limited data settings is analyzed in \cite{chai2022awn}. The analysis findings indicate that ANP becomes ineffective when fewer than 100 data samples are available (see \cite[Fig. 1]{chai2022awn}). However, this analysis only considers CIFAR-10. Therefore, the 100-sample threshold reported in \cite{chai2022awn} cannot be considered universally applicable.

To overcome the impact of limited data on the performance of weight masking, \cite{chai2022awn} propose applying perturbations to the input space rather than to $\theta$. They redesign the inner maximization problem to identify an input perturbation $\delta$ that maximizes classification error when applied to $\mathcal{D}_{m}$. Given $\widetilde{\mathcal{D}}_{m} = \{(\tilde{x}, y) \: | \: \tilde{x} = x + \delta \: | \: (x, y) \in \mathcal{D}_{m}\}$, the inner maximization problem is expressed as

\begin{equation} \label{awn-max}
    \max_{\|\delta\|_{1} \leq \epsilon} \mathcal{L}_\mathrm{CE}(\widetilde{\mathcal{D}}_{m}, \theta),
\end{equation}
where $\| \delta \|_{1}$ is bound by $\epsilon$. Notably, the same $\delta$ is applied to all elements of $\mathcal{D}_{m}$. Thus, \cite{chai2022awn} solve the following minimax optimization problem

\begin{equation} \label{awn-min}
\begin{split}
\min_{\mathbf{m} \in [0, 1]} \biggl\{& \lambda_{1} \mathcal{L}_\mathrm{CE}(\mathcal{D}_{m}, \mathbf{m} \space \odot \theta)\\
&+\lambda_{2} \max_{\|\delta\|_{1} \leq \epsilon}\left[ \mathcal{L}_\mathrm{CE}(\widetilde{\mathcal{D}}_{m}, \mathbf{m} \space \odot \theta) \right] + \lambda_{3} \|\mathbf{m} \|_{1} \biggr\},
\end{split}
\end{equation}
where $\lambda_{1}$, $\lambda_{2}$, and $\lambda_{3}$ are hyperparameters that balance the influence of the three loss terms, and $\|\mathbf{m} \|_{1}$ serves an additional regularization term to encourage sparsity. Note that \cite{chai2022awn} omit the final binarization step employed in ANP, retaining $\mathbf{m}$ as a soft mask. 

Promoting sparsity in $\mathbf{m}$ leads to significant pruning of $\theta$, which is expected to result in low bias and high variance given $\mathcal{D}_{m}$. Moreover, this additional term is not normalised to account for the size of $\mathbf{m}$. Therefore, identifying optimal values for $\lambda_{1}$, $\lambda_{2}$, and $\lambda_{3}$ that yield consistent performance across different model architectures is challenging.

\subsubsection{RNP}

Distinct from ANP~\cite{wu2021anp} and AWN~\cite{chai2022awn}, \cite{li2023rnp} introduce a unique unlearning strategy. Rather than learning $\xi$ or $\delta$, \cite{li2023rnp} first \textit{unlearn} the clean task by solving the following optimization problem

\begin{equation}
    \max_{\theta} \mathcal{L}_\mathrm{CE}(\mathcal{D}_{m}, \theta).
\end{equation}
The resulting set of unlearned parameters is denoted as $\hat{\theta}$. In \cite{li2023rnp}, it is argued that this process leads to unlearning of the model's clean components while preserving the backdoor. Moreover, they suggest that the unlearned model exhibits biased misclassification toward the backdoor target. While a set of feature maps are presented in support of this in \cite[Fig. 2]{li2023rnp}, this conclusion is not validated across a diverse range of settings.
Using $\hat{\theta}$, \cite{li2023rnp} then proceed to learn $\mathbf{m}$ by solving the following optimization problem  
\begin{equation}
    \min_{\mathbf{m} \in [0, 1]} \mathcal{L}_\mathrm{CE}(\mathcal{D}_{m}, \mathbf{m} \space \odot \hat{\theta}).
\end{equation}
The authors of \cite{li2023rnp} assert that this recovery procedure can differentiate between clean and backdoor filters, given that $\hat{\theta}$ exhibits biased misclassification towards the backdoor target. Specifically, they suggest that this procedure removes backdoor filters by setting their corresponding elements in $\mathbf{m}$ to $0$. Importantly, once an optimal solution to $\mathbf{m}$ is obtained, it is applied to the original parameters, $\theta$. In a manner similar to ANP, \cite{li2023rnp} binarize $\mathbf{m}$ using a threshold value or a fixed pruning ratio. 

\subsection{Additive Pruning}

The pruning methods discussed thus far either directly prune or mask existing model components. In contrast, \cite{wang2023mm} and \cite{zhu2024npd} introduce learning additional model components that integrate with the existing structure. These additional components function as quasi-filters, targeting the removal of backdoor tasks through a mechanism akin to pruning. Consequently, we categorize these approaches loosely under the umbrella of pruning.

\subsubsection{MM-BD}

In \cite{wang2023mm}, the task of backdoor mitigation is framed as a bounding problem. Initially, \cite{wang2023mm} observe that backdoor samples tend to trigger large activations, leading to unusually high decision-making confidence (see \cite[Fig. 9]{wang2023mm}). To quantify this difference in confidence, \cite{wang2023mm} calculate the maximum margin statistic of $x$ given $y$ as 

\begin{equation}
    \mathcal{G}(x, y, \theta) = s_{y}(a) - \max_{k \in \mathcal{M} \setminus y} s_{k}(a),
\end{equation}
where $s_n$ selects the $n$th logit from the softmax output $a$ given $x$, and $\mathcal{M}$ represents the set of possible labels. Their analysis reveals that backdoor samples exhibit significantly larger confidence compared to clean samples (see \cite[Fig. 4(a)]{wang2023mm}). The authors conjecture that this is due to the abnormally large activations influencing the model's decision-making. 

To counteract the impact of these large activations have on decision-making, \cite{wang2023mm} propose learning a set of upper-bound values $\mathbf{B} = \{b_1, \dots, b_{L}\}$ for each non-linear activation layer, such as ReLU. These bounds are learned in a channel-wise manner and used during the forward pass to constrain activation range within each channel. To learn $\mathbf{B}$, the following optimization problem is solved

\begin{equation}
    \min_{\mathbf{B}} \frac{1}{|\mathcal{D}_{m}| \times |\mathcal{M}|} \sum_{(x, y) \in \mathcal{D}_{m}} [f(x, \theta_{\mathbf{B}}) - f(x, \theta)]^{2} + \lambda \| \mathbf{B} \|_{2},
\end{equation}  
where $\theta_{\mathbf{B}}$ represents the model parameters $\theta$ combined with the learned bounding values $\mathbf{B}$. The objective is to find the bounding values $\mathbf{B}$ that minimally impact the classification of $\mathcal{D}_{m}$.

Similar to AWN, $\| \mathbf{B} \|_{2}$ is not normalised to account for variations in model architectures. In addition, this term does not adjust for differences in the scale of activation values across layers. Variations in activation scale are likely to directly affect the values of $\mathbf{B}$ and influence the choice of $\lambda$. Another critical assumption made in \cite{wang2023mm} is that setting an upper bound for each channel is sufficient to restore correct classification of $\mathcal{D}_{b}$. However, \cite{wang2023mm} does not report whether their mitigation strategy successfully restores the correct classification of $\mathcal{D}_{b}$ (i.e., $\hat{x}$ is classified as $y$). Additionally, \cite{wang2023mm} does not demonstrate the effect of $\mathbf{B}$ on the channel-wise activation of $\mathcal{D}_{b}$ across layers. This information is crucial for understanding how much of the activation map is clipped to the bounding value, thereby indicating the effectiveness of clipping large activations as a mitigation strategy.      
 
\subsubsection{NPD} \label{method:npd}

Unlike MM-BD~\cite{wang2023mm}, \cite{zhu2024npd} approaches backdoor mitigation from a more traditional pruning perspective. To implement pruning, \cite{zhu2024npd} introduces a 1x1 convolutional layer into the model, suggesting that this layer is added close to the final layer of the feature extraction. Referred to as a polarizer, this layer learns a set of parameters $\boldsymbol{w}$ that filter out channels possibly associated with the backdoor task. The 1x1 convolutional layer maintains the same number of channels as the preceding layer, scaling each output channel by the corresponding value in $\boldsymbol{w}$. The augmented model $f_{w, \theta}$ is parameterized by $\theta$ and $\mathbf{w}$. However, $\theta$ remains fixed throughout, and therefore is excluded from subsequent equations.

Similar to AWN~\cite{chai2022awn}, \cite{zhu2024npd} first approximates the backdoor trigger $\rho$ as an input perturbation $\delta$. However, unlike AWN, \cite{zhu2024npd} model the trigger distribution in a sample-specific manner, learning a distinct value of $\delta$ for each $(x, y) \in \mathcal{D}_{m}$. Given $x$, $\delta$ is learned by solving the following optimization problem

\begin{equation}
    \min_{\|\delta\|_{p} \leq \epsilon} \mathcal{L}_\mathrm{CE}(x + \delta, \tilde{y}, \mathbf{w}),
\end{equation}
where $\tilde{y}$ is an estimate of the target class, for which \cite{zhu2024npd} provide several heuristics. Under the assumption that the defender does not know the backdoor target, \cite{zhu2024npd} suggest using the second-largest logit for $x$.However, \cite{zhu2024npd} does not quantitatively validate how frequently the second-largest logit corresponds to the backdoor target. Using $\widetilde{\mathcal{D}}_{m} = \{(\tilde{x}, \tilde{y}, x, y) \: | \: \tilde{x} = x + \delta \: | \: (x, y) \in \mathcal{D}_{m}\}$, they solve the following optimization problem

\begin{equation}
    \mathcal{L}_{\mathrm{ASR}}(\tilde{x}, \tilde{y}, \mathbf{w}) = -\log(1-s_{\tilde{y}}(\tilde{a})),
\end{equation}

\begin{equation}
\begin{split}
    \mathcal{L}_{\mathrm{BCE}}(\tilde{x}, y, \mathbf{w}) = -\log(s_{y}(\tilde{a})) - \log(1-\max_{k\neq y}s_{k}(\tilde{a})),
\end{split}
\end{equation}

\begin{equation}
\begin{split}
\min_{\mathbf{w}} \biggl\{& \frac{1}{|\widetilde{\mathcal{D}}_{m}|} \sum_{(\tilde{x}, \tilde{y}, x, y) \in \widetilde{\mathcal{D}}_{m}} \lambda_{1} \mathcal{L}_\mathrm{CE}(x, y, \mathbf{w})\\
&+\lambda_{2} \mathcal{L}_\mathrm{ASR}(\tilde{x}, \tilde{y}, \mathbf{w}) + \lambda_{3} \mathcal{L}_\mathrm{BCE}(\tilde{x}, y, \mathbf{w})\biggr\},
\end{split}
\end{equation}
where $\tilde{a}=f(\tilde{x}, \theta)$, and $\lambda_{1}$, $\lambda_{2}$ and $\lambda_{3}$ are hyperparameters that control the influence of each term. This design optimizes $\mathbf{w}$ to alleviate the impact of $\delta$ when applied to $\mathcal{D}_{m}$. The term $\mathcal{L}_\mathrm{ASR}$ penalizes the classification of $\tilde{x}$ as $\tilde{y}$, while $\mathcal{L}_\mathrm{BCE}$, similar to the margin statistic proposed in \cite{wang2023mm}, encourages confident classification of $\tilde{x}$ as $y$, its correct label.

\subsection{Summary}

Model pruning is an important strategy in mitigating backdoor attacks, building on the hypothesis that neural networks can be decomposed into distinct components responsible for either clean or backdoor tasks. By selectively pruning the components related to backdoor behaviors, researchers aim to restore model integrity while preserving performance on the original task.

Metric-based pruning approaches quantify each model component's contribution to backdoor behavior through various metrics. Approaches like FP remove filters based on their activation patterns, while BNP utilizes distribution statistics to identify backdoor components. CLP's Lipschitz-based approach uses approximations of the channel Lipschitz constants to guide pruning decisions. Masking-based pruning techniques optimize a parameter mask to remove backdoor functionality. Approaches such as ANP, AWM, and RNP employ optimization frameworks to iteratively refine the mask, targeting backdoor components while retaining model accuracy. Lastly, additive pruning approaches introduce new components into the network, functioning as filters that mitigate backdoor influences without directly removing existing filters. These approaches exemplify innovative ways of addressing backdoor mitigation while maintaining the overall structure of the model. Each pruning approach offers distinct advantages and limitations that require robust evaluation across a broad range of settings.

\section{Fine Tuning} \label{fine-tuning}

An alternative approach to model pruning for backdoor mitigation is fine-tuning. Instead of removing a subset of $\theta$, fine-tuning methods adjust the values of $\theta$ to eliminate the backdoor. A key aspect of these methods is the objective function, which typically incorporates one or more carefully designed regularisation terms. These terms are usually selected based on specific insights gained from preliminary investigations. In this section, we review the most prominent fine-tuning approaches, categorized into distinct subgroups based on their unique methodologies. 

\subsection{Conventional Fine-Tuning}

The first subcategory of fine-tuning methods follows a more conventional approach. Here, conventional refers to the proposed optimization problem being closely aligned with~\eqref{general-training-objective}. 

\subsubsection{FST}

In their preliminary investigation, \cite{min2024fst} evaluate the effectiveness of minimisation~\eqref{general-training-objective} given $\mathcal{D}_{m}$ for mitigating backdoor tasks. Their analysis decomposes $\theta$ into $\varphi$, parameters associated with the feature extractor, and $\omega$, parameters associated with the linear classifier. They tested fine-tuning various combinations of $\varphi$ and $\omega$, concluding that the minimisation~\eqref{general-training-objective} for any combination of these parameters was largely ineffective in removing backdoors (see \cite[Table 1]{min2024fst}).  

Building on these findings, \cite{min2024fst} propose reinitializing $\omega$ by assigning new random values to $\omega$ before jointly fine-tuning both $\varphi$ and $\omega$. To enhance this process, they introduce a regularisation term that encourages divergence between the original and new values of $\omega$, referred to as $\hat{\omega}$ and $\omega$, respectively. This leads to minimizing the following objective function

\begin{equation}
    \min_{\theta} \mathcal{L}_{\mathrm{CE}}(\mathcal{D}_{m}, \theta) + \lambda \omega^\intercal \hat{\omega}, \quad \text{s.t.} \; \|\omega\|_{2} = \|\hat{\omega}\|_{2},
\end{equation}  
where $\lambda$ is a hyperparameter controlling the influence of the regularization term. According to \cite{min2024fst}, $\omega^\intercal \hat{\omega}$ discourages $\omega$ from learning the same relationships between the penultimate set of features and class labels. The constraint $\|\omega\|_{2} = \|\hat{\omega}\|_{2}$ is applied to minimise the impact of $\omega^\intercal \hat{\omega}$ on the overall loss during fine-tuning. While this additional term improves the unnormalised loss components used by AWN and MM-BD, it does not explicitly account for differences in the size of $\omega$ used by each model architecture (i.e., the number of neurons in dense layers).

\subsubsection{FT-SAM}

In addition, \cite{zhu2023ft-sam} investigate the effectiveness of traditional fine-tuning in mitigating backdoors. Similar to FST~\cite{min2024fst}, they find it ineffective. Their analysis of the $\omega$ norms, the magnitudes of each parameter, revealed minimal changes following fine-tuning (see \cite[Fig. 2]{zhu2023ft-sam}). They hypothesise that fine-tuning fails because it is unable to escape the local minima it finds, allowing the backdoor to persist. Additionally, \cite{zhu2023ft-sam} demonstrates that $\omega$ norms are positively correlated with TAC [see~\eqref{clp:tac}], a metric previously used in CLP~\cite{zheng2022clp} to quantify activation difference between $\mathcal{D}_{c}$ and $\mathcal{D}_{b}$. 

Inspired by sharpness-aware minimisation (SAM) methods \cite{foret2020sharpness}, \cite{zhu2023ft-sam} propose a minimax optimisation method designed to escape sharp local minima. Similar to ANP~\cite{wu2021anp}, the inner minimisation seeks an $\ell_2$-bounded perturbation $\xi$ that maximises the classification loss for $\mathcal{D}_{m}$. However, rather than identifying a weight mask $\mathbf{m}$, they fine-tune $\theta$ directly in the outer minimisation step. This leads to solving the following optimization problem

\begin{equation}
\begin{split}
    \min _{\theta} \max _{\left\|\mathbf{T}_{\theta}^{-1} \xi\right\|_{2} \leq \epsilon} \mathcal{L}_{\mathrm{CE}}(\mathcal{D}_{m}, \theta+\xi),
\end{split}
\end{equation}
where $\mathbf{T}_{\theta}=\operatorname{diag}\left(\left|\theta_{1}\right|,\ldots,\left|\theta_{L}\right|\right)$, with $\theta_{i}$ being the $i^{\text{th}}$ parameter in $\theta$ and $\epsilon$ serving as a hyperparameter controlling the perturbation budget. The traditional $\ell_2$-bounding constraint is modified to $\|\mathbf{T}_{\theta}^{-1} \xi\|_{2} \leq \epsilon$, allowing larger perturbations to be applied to elements of $\theta$ with larger norms, as their corresponding component in $\mathbf{T}_{\theta}^{-1}$ approaches zero. This evaluates the stability of $\theta$, quantified as $\mathcal{L}_{\mathrm{CE}}$ on $\mathcal{D}_{m}$, when perturbed by $\xi$. 

The perturbation constraint $\|\mathbf{T}_{\theta}^{-1} \xi\|_{2} \leq \epsilon$ used by \cite{zhu2023ft-sam} is applied during each gradient decent step. This implies that $\theta$ can drift significantly from its initial values after several steps. Moreover, since the estimation of $\xi$ relies on $\mathcal{D}_{m}$, it is prone to having low bias and high variance.


\subsection{Knowledge Distillation}

Inspired by its success in other learning settings, two proposals explore how knowledge distillation (KD) can be leveraged for backdoor mitigation. Traditionally used to \textit{transfer knowledge} from larger to smaller models, KD has been effectively applied in tasks such as image classification \cite{gou2021knowledge}. In the context of backdoor mitigation, the distillation process is reframed as a \textit{knowledge filtering} task. Instead of transferring all knowledge from the original model, the goal of \textit{knowledge filtering} is to distill only the information relevant to the clean task, thereby eliminating the backdoor-related information.

A common approach to KD involves a teacher-student architecture. In typical applications, the teacher model is a larger, more capable model whose knowledge is transferred to a smaller student model. However, in the case of backdoor mitigation, access to a non-backdoored teacher model is not possible. Subsequently, approaches that employ this method must overcome this challenge.   

\subsubsection{NAD} 

To implement KD, \cite{li2021nad} introduce a new attention-based method. Rather than relying on feature maps (i.e., intermediate activation outputs) to perform KD, \cite{li2021nad} suggest using attention maps. These maps compress the channel dimension of the feature maps, utilizing an attention operator $\mathcal{A}$, which maps from $\mathbb{R}^{c \times h \times w}$ to $\mathbb{R}^{h \times w}$. They propose the following two variants

\begin{equation}
    \mathcal{A}_{\mathrm{sum}}^{p}(x, \theta, l) = \sum_{i=1}^{c} |f_{l,i}(x,\theta)|^{p}
\end{equation}
\begin{equation}
    \mathcal{A}_{\mathrm{mean}}^{p}(x, \theta, l) = \frac{1}{c} \sum_{i=1}^{c} |f_{l,i}(x,\theta)|^{p}
\end{equation}
where $f_{l,i}$ is the $i^{\text{th}}$ channel activation of $x$ at the $l^{\text{th}}$ layer and $p > 1$. To perform KD within the teacher-student framework, \cite{li2021nad} first fine-tune $\theta$ using $\mathcal{D}_{m}$ resulting in a teacher model with parameters $\theta_{T}$. The student's parameters, $\theta_{S}$, are the original parameters $\theta$ that have not been fine-tuned. The KD is then performed by comparing the activation maps between the teacher and student models. To achieve this, \cite{li2021nad} use $\mathcal{A}_{sum}^{p}$ to define a distillation loss as

\begin{equation}
    \mathcal{L}_{\mathrm{NAD}}(x, \theta_{T}, \theta_{S}, l) = \left\| \frac{\mathcal{A}(x, \theta_{T}, l)}{\|\mathcal{A}(x, \theta_{T}, l)\|_{2}} - \frac{\mathcal{A}(x, \theta_{S}, l)}{\|\mathcal{A}(x, \theta_{S}, l)\|_{2}}\right\|_{2},
\end{equation}
and solve the following optimization problem
\begin{equation}
    \min_{\theta_{S}} \mathcal{L}_\mathrm{CE}(\mathcal{D}_{m}, \theta_{S}) + \frac{\lambda}{|\mathcal{D}_{m}|} \sum_{(x, y) \in \mathcal{D}_{m}} \sum_{l=1}^{L} \mathcal{L}_\mathrm{NAD}(x, \theta_{T}, \theta_{S},l),
\end{equation}
where $\lambda$ controls the contribution of the distillation loss. According to \cite{li2021nad}, the inclusion of $\mathcal{L}_{\mathrm{NAD}}$ helps regularise $\theta$ by aligning the activation maps of $\theta_{S}$ and $\theta_{T}$ thus removing the backdoor behaviour. However, since knowledge is distilled from a fine-tuned version of $\theta$, the effectiveness of this approach is unclear if fine-tuning does not successfully remove the backdoor. This concern was underscored by the initial findings of both FST and FT-SAM, where fine-tuning alone proved ineffective at eliminating the backdoor.

\subsubsection{BCU} 

In contrast to NAD~\cite{li2021nad}, \cite{pang2023bcu} propose using softmax probabilities $a$ of $x$ to perform KD rather than attention maps. Specifically, \cite{pang2023bcu} compare the temperature-scaled softmax probability scores $\tilde{a}_{T}=f(x, \theta_{T})$ and $\tilde{a}_{S}=f(x, \theta_{S})$, produced by $\theta_{S}$ and $\theta_{T}$, respectively, to facilitate KD. To compare $\tilde{a}_{T}$ and $\tilde{a}_{S}$, \cite{pang2023bcu} employ KL-Divergence $\mathcal{L}_{\mathrm{KL}}$ and solve the following optimization problem

\begin{equation}
    \min_{\theta_{S}} \mathcal{L}_{\text{KL}}(\tilde{a}_{T}, \tilde{a}_{S}).
\end{equation}

Rather than fine-tuning $\theta$ to produce $\theta_{T}$, \cite{pang2023bcu} reinitialise a subset of $\theta_{S}$, note that, $\theta_{T} = \theta$. Unlike FST~\cite{min2024fst}, \cite{pang2023bcu} uniformly reinitialize $n$ proportion ($0 \leq n \leq 1$) of the parameters within each layer, with $n$ increasing for deeper layers. They identified this reinitialization strategy as the best approach through a series of experiments. However, \cite{pang2023bcu} assert that the reinitialization step significantly impairs the model's ability to perform both clean and backdoor tasks. Consequently, when minimizing the proposed objective function, only the knowledge related to the clean task is transferred from the teacher to the student, severing the link between the trigger pattern and the backdoor task. 

Unlike previous proposals, the use of KL-divergence by \cite{pang2023bcu} makes their approach dataset-agnostic. As a result, a defender can use any labeled or unlabeled in- or out-of-distribution dataset compatible with their model (i.e., having the same input dimensionality) to perform backdoor mitigation.

\subsection{Synthesis Unlearn}

The fine-tuning approaches discussed thus far utilise $\mathcal{D}_{m}$ to fine-tune $\theta$. An alternative strategy involves synthesizing a set of surrogate backdoor data $(\tilde{x}, \tilde{y}) \in \widetilde{\mathcal{D}}_{m}$, which is used alongside $\mathcal{D}_{m}$ for fine-tuning. Here, $\tilde{x}$ and $\tilde{y}$ represent the surrogate backdoor data. The inclusion of this initial synthesis step allows these approaches to exploit information from $\widetilde{\mathcal{D}}_{m}$ to directly unlearn the backdoor task.

\subsubsection{MESA}

To synthesize $\tilde{x}$, \cite{qiao2019mesa} propose training a generative model $G$, parameterized by $\gamma$, to replicate the trigger distribution used by the adversary. To train $G$, they introduce a new maximum-entropy staircase approximation algorithm. This algorithm trains $G$ as a combination of $n$ sub-models that collectively generate a candidate trigger for a given input $x$. However, using a set of sub-models to train $G$ requires the defender to know the trigger's position, approximate size, and the backdoor target. The optimization problem they solve is

\begin{equation}
\begin{aligned}
    \min_{\theta} & \frac{1}{|\mathcal{D}_{m}|} \sum_{x, y \in \mathcal{D}_{m}} \left[\lambda\mathcal{L}_{\mathrm{CE}}(x, y, \theta)+(1-\lambda)\mathcal{L}_{\mathrm{CE}}(\tilde{x}, y, \theta)\right]
\end{aligned}
\end{equation}
where $\tilde{x} = x + G(\gamma)$ and $\lambda \in [0, 1]$ is a hyperparameter selected by the defender to control how many elements of $\mathcal{D}_{m}$ have $\delta$ applied. This approach aims to balance restoring the classification of $\tilde{x}$ to $y$ while preserving the original classification performance.

\subsubsection{BAERASER}

Inspired by MESA~\cite{qiao2019mesa}, \cite{liu2022baeraser} adopt the same synthesis method for generating surrogate backdoor data but proposes a different fine-tuning step. Using $G$, they generate a surrogate backdoor dataset $\widetilde{\mathcal{D}}_{m}$ that is used in conjunction with $\mathcal{D}_{m}$ to unlearn the backdoor task. The surrogate dataset is defined as $\widetilde{\mathcal{D}}_{m} = \{(\tilde{x}, \tilde{y}) \: | \: \tilde{x} = x + \delta \: | \: \delta = G(x, \gamma) \: | \: (x, y) \in \mathcal{D}_{m}\}$, assuming the defender has access to $\tilde{y}$. To perform unlearning, \cite{liu2022baeraser} solve the following optimization problem

\begin{equation}
\begin{split}
    \min_{\theta} \lambda_{1} [\mathcal{L}_{\mathrm{CE}}(\mathcal{D}_{m}, \theta) - \mathcal{L}_{\mathrm{CE}}(\widetilde{\mathcal{D}}_{m}, \theta)] + \lambda_{2} \sum_{l=1}^{L} w_{l} \|\theta_{l} - \bar{\theta}_{l}\|_{1},
\end{split}
\end{equation}
where $\lambda_{1}$ and $\lambda_{2}$ control the strength of the two loss terms. The first term encourages misclassification of $\widetilde{\mathcal{D}}_{m}$ by subtracting its loss from that of $\mathcal{D}_{m}$. However, this term is unbounded and can dominate the optimization after a few iterations. The second loss term regularizes the solution by minimising the layer-wise distance between $\theta$ and the original value $\bar{\theta}$, using a layer-wise scalar weight $w_{l}$. 

\subsubsection{NC}

Unlike both MESA and BAEARSER, \cite{wang2019nc} propose a method that removes the assumption of knowing the backdoor target and the approximate size and position of $\rho$. To achieve this, \cite{wang2019nc} learn an input perturbation $\delta$ that replaces specific image pixels using a binary mask $\mathbf{m}$. Here, $\delta$ and $\mathbf{m}$ are 3D and 2D matrices, respectively, with width and height dimensions matching $x$. To apply $\delta$ to $x$ given $\mathbf{m}$, the function $A(x, \mathbf{m}, \delta) \rightarrow \hat{x}$ is utilized where if $\mathbf{m}_{j,i} = 1$, $A$ replaces the pixel in the $j^{\text{th}}$ row and $i^{\text{th}}$ column of $x$ with the corresponding value in $\delta$. To learn $\delta$ and $\mathbf{m}$, \cite{wang2019nc} solve the following optimization problem

\begin{equation}
    \min_{\mathbf{m}, \delta} \sum_{x \in \mathcal{D}_{m}} \mathcal{L}_{\mathrm{CE}}(A(x, \mathbf{m}, \delta), t) + \lambda \|\mathbf{m}\|_{1},
\end{equation}
where $\lambda$ controls the strength of the second regularisation term, which promotes sparsity in the solution for $\mathbf{m}$. Since $t$ is unknown to the defender, a unique solution for $\delta$ and $\mathbf{m}$ is determined separately for each class. An anomaly detection mechanism, using median absolute deviation, is then employed to identify anomalous class pairs. If such a pair is found, the proposed approach prunes the final dense layer of the model to mitigate the effect of the backdoor. To perform model pruning, the TAC metric [cf.~\eqref{clp:tac}] is used. Neurons that exhibit the largest average activation difference when $\delta$ is applied to $D_{m}$ using $A$ are iteratively pruned. Despite relying on model pruning, the method shares key similarities with MESA and BAEARSER, making it relevant to this section.

\subsection{Adversarial Training}

Instead of approximating $\delta$ in a discrete step, recent works have incorporated concepts from adversarial training to perform mitigation. In essence, these approaches alternate between an adversarial objective and a mitigation objective. However, the critical distinction lies in the design of the adversarial objective. Unlike traditional adversarial examples, the adversarial objective is specifically tailored to generate surrogate backdoor images. This ensures that the outer mitigation objective remains effective, allowing the model to unlearn the backdoor task while preserving performance on the clean task. 

\subsubsection{PBE} \label{method:pbe}

In \cite{mu2023pbe}, the authors explore the behaviour of untargeted adversarial attacks on backdoored models. Given $x$ and $y$, they generate an input perturbation $\delta$ by solving the following adversarial optimization problem

\begin{equation}
    \max_{\|\delta\|_{2} \leq \epsilon} \mathcal{L}_\mathrm{CE}(\tilde{x}, y, \theta),
\end{equation}
where $\epsilon$ controls the strength of the perturbation and $\tilde{x} = x + \delta$. Upon analyzing the classification of $\tilde{x}$, \cite{mu2023pbe} observed that a backdoored model tends to classify $\tilde{x}$ as the backdoor target, whereas a benign model produces a uniform distribution (see \cite[Fig. 4]{mu2023pbe}). Hence, they hypothesised that $\tilde{x}$ interacts with the backdoored model similarly to $\hat{x}$, the actual backdoor version of $x$. However, it is important to note that the proportion of samples classified as the target class in \cite[Fig. 4]{mu2023pbe} does not exceed 61\%.

To exploit this observation, \cite{mu2023pbe} propose a fine-tuning strategy where $\theta$ is trained using both $\mathcal{D}_{m}$ and $\widetilde{\mathcal{D}}_{m} = \{(\tilde{x}, y) \: | \: \tilde{x} = x + \delta \: | \: (x, y) \in \mathcal{D}_{m}\}$. They alternate between solving the following two optimization problems

\begin{equation}
    \min_{\theta} \mathcal{L}_\mathrm{CE}(\mathcal{D}_{c}, \theta), \quad \min_{\theta} \mathcal{L}_\mathrm{CE}(\widetilde{\mathcal{D}}_{m}, \theta),
\end{equation}
where $\delta$ is computed using the PGD attack \cite{madry2018pgd}.

\subsubsection{i-BAU}

In \cite{zeng2022i-bau}, the authors propose a redesigned adversarial objective aimed at identifying a universal input perturbation $\delta$, which functions similarly to AWN~\cite{chai2022awn}. Here, universal refers to a perturbation that applies to all elements within $\mathcal{D}_{m}$. Using $\widetilde{\mathcal{D}}_{m} = \{(\tilde{x}, y) \: | \: \tilde{x} = x + \delta \: | \: (x, y) \in \mathcal{D}_{m}\}$, \cite{zeng2022i-bau} set up the following minimax optimization problem

\begin{equation} \label{i-bau:adv-learning}
    \min_{\theta} \max_{\|\delta\|_{2} \leq \epsilon} \mathcal{L}_\mathrm{CE}(\widetilde{\mathcal{D}}_{m}, \theta).
\end{equation}
However, their experiments (see~\cite[Fig. 1]{zeng2022i-bau}) reveal that solving this minimax problem directly often yields unstable and unreliable results. This instability is attributed to the inner maximisation step failing to find an optimal solution for $\delta$. To alleviate this issue, \cite{zeng2022i-bau} propose solving the outer minimisation step using the following gradient

\begin{equation} \label{i-bau:gradient}
\begin{split}
    \nabla_{\theta}\mathcal{L}_{\mathrm{CE}}(\widetilde{\mathcal{D}}_{m}, \theta) +  (\nabla \delta)^{T} \nabla_{\delta}\mathcal{L}_{\mathrm{CE}}(\widetilde{\mathcal{D}}_{m}, \theta),
\end{split}
\end{equation}
where $\nabla \delta$ is the \textit{response Jacobian} of the inner maximisation problem. Given that the inner maximisation step produces a suboptimal solution for $\delta$, they calculate $\nabla \delta$ as 

\begin{equation}
    \nabla \delta = -\left(\nabla_{\delta}^{2}\mathcal{L}_{\mathrm{CE}}(\widetilde{\mathcal{D}}_{m}, \theta)\right)^{-1} \nabla_{\delta,\theta}^{2}\mathcal{L}_{\mathrm{CE}}(\widetilde{\mathcal{D}}_{m}, \theta).
\end{equation}
This ensures that the response Jacobian captures the sensitivity of $\delta$ to changes in $\theta$, while $\nabla_{\delta}\mathcal{L}_{\mathrm{CE}}(\widetilde{\mathcal{D}}_{m}, \theta)$ captures the direct sensitivity of $\delta$. These adjustments allow the gradient update for $\theta$ to incorporate the sensitivity of $\delta$, resulting in a more stable and reliable solution for adversarial fine-tuning. The complexity of the gradient estimation method in~\cite{zeng2022i-bau} increases the likelihood of overfitting, as $\nabla_{\delta}$ is dependent on the estimation of $\nabla \delta$. Notably, $\nabla \delta$ requires estimation using second-order algorithms. While \cite{zeng2022i-bau} asserts that these methods are robust to inaccuracies in the Hessian, the referenced literature assumes access to a large training dataset.

\subsubsection{SAU}

To enhance existing approaches, \cite{wei2023sau} propose filtering candidate perturbations $\delta$ based on their ability to induce consistent misclassification across two classifiers. In this context, consistent misclassification means that both classifiers classifying $\tilde{x}$ as the same incorrect class $\tilde{y}$. Formally, \cite{wei2023sau} optimize $\delta$ for each $(x,y) \in \mathcal{D}_{m}$ by solving the following optimization problem

\begin{equation} \label{sau:generation-method}
\begin{split}
\max_{\|\delta\|_{p} \leq \epsilon} \biggl\{ \frac{\lambda_{1}}{2} [\mathcal{L}_{\mathrm{CE}}(\tilde{x}, y, \theta) &+ \mathcal{L}_{\mathrm{CE}}(\tilde{x}, y, \bar{\theta})] \\
&- \lambda_{2} \; \mathrm{JS}(f(\tilde{x}, \theta), f(\tilde{x}, \bar{\theta}))\biggr\},
\end{split}
\end{equation}
where $\mathrm{JS}$ is the Jensen-Shannon divergence, and $\lambda_{1}$ and $\lambda_{2}$ are hyperparameters. Since the defender does not have access to two classifiers, the original model parameters $\bar{\theta}$ are used as the second classifier, $\bar{\theta}$ remaining fixed. Unlike PBE and NPD, SAU aims to distinguish between adversarial examples and backdoor triggers. To achieve this, they ensure that $\delta$ causes both $\theta$ and $\bar{\theta}$ to misclassify $\tilde{x}$ consistently, a behaviour more characteristic of backdoor triggers than typical adversarial examples (see~\cite[Fig. 2]{wei2023sau}).

Once the surrogate backdoor dataset $\widetilde{\mathcal{D}}_{m} = \{(\tilde{x}, x, \tilde{y}, y) \: | \: \tilde{y} \leftarrow f(\tilde{x}, \theta) \: | \: \: \tilde{x} = x + \delta \: | \: (x, y) \in \mathcal{D}_{m}\}$ is generated, \cite{wei2023sau} solve the following optimization problem to fine-tuned the model
\begin{equation}
\begin{split}
\min_{\theta}\biggl\{ \frac{1}{|\widetilde{\mathcal{D}}_{m}|} &\sum_{(\tilde{x}, x, \tilde{y}, y) \in \widetilde{\mathcal{D}}_{m}} \lambda_{3} \; \mathcal{L}_{\mathrm{CE}}(x, y, \theta) \\
&- I(\tilde{y} \neq y) \log [1 - s_{\tilde{y}}(f(\tilde{x}, \theta))]\biggr\},
\end{split}
\end{equation}
where $I$ is the indicator function that is 1 if $\tilde{x}$ is misclassified and $\lambda_{3}$ is another hyperparameter. This formulation balances the performance on $\mathcal{D}_{m}$, represented by the first term, with correcting the classification of $\tilde{x}$ to $y$, captured by the second term.

\subsection{Summary}

Fine-tuning, as an alternative to model pruning for backdoor mitigation, adjusts the parameters of a model rather than removing them. This strategy is typically governed by an objective function incorporating tailored regularization terms. Conventional fine-tuning approaches, such as FST and FT-SAM, attempt to adjust model weights to eliminate backdoors, though they often struggle with escaping local minima and thus fail to fully mitigate the backdoor threat. More advanced approaches based on KD, such as NAD and BCU, reframe fine-tuning as a process of filtering out harmful information. These approaches leverage the distillation of knowledge from a teacher model to a student model to effectively mitigate backdoor attacks while retaining the model's performance on clean data. Additionally, approaches such as MESA and BAERASER employ surrogate data generation to support the fine-tuning process. However, these approaches, rely on the defender having prior knowledge about the specific trigger used by an adversary. To remove this assumption, approaches such as PBE, i-BAU, and SAU modify existing adversarial training techniques. Similar to pruning approaches, each fine-tuning approach presents unique strengths and weaknesses, necessitating thorough evaluation across a diverse range of settings to fully understand their effectiveness.

\section{Experimental Setup}

In this section, we describe the experimental setup of our extensive evaluations covering 16 of the 18 approaches discussed in Sections~\ref{model-pruning} and ~\ref{fine-tuning}. We benchmark each approach across a wide variety of settings as our evaluations span various backdoor attacks, model architectures, datasets, and poisoning ratios, resulting in a total of 288 distinct attack scenarios. Moreover, unlike \cite{wu2022backdoorbench}, we test each considered mitigation approach across three data availability settings, leading to 122,236 individual experiments in total. For our evaluations, we employ the \textit{BackdoorBench} toolkit~\cite{wu2022backdoorbench}, as it provided most of the required functionality. However, we have made several key modifications to this toolkit, such as incorporating the implementation of five additional mitigation approaches.


\subsection{Attacks}

Our evaluations include all backdoor attacks introduced in section~\ref{prelim:backdoor-attacks}, with the exception of LIRA. We exclude LIRA due to its poor performance during an initial set of experiments. Therefore, the attacks we consider are BadNets~\cite{gu2017badnets}, Blended~\cite{chen2017targeted}, Signal~\cite{barni2019new}, LF~\cite{zeng2021rethinking}, SSBA~\cite{li2021invisible}, IAB~\cite{nguyen2020input}, BPP~\cite{wang2022bppattack}, and WaNet~\cite{nguyen2021wanet}. For each attack, we use the default configurations provided in \textit{BackdoorBench}. We implement the attacks using poisoning ratios of 1\%, 5\%, and 10\%, selected based on the findings reported in~\cite{wu2022backdoorbench}. 

\subsection{Mitigation Methods}

With the exception of MESA, BAERASER and BCU, all approaches discussed in sections~\ref{model-pruning} and \ref{fine-tuning} are evaluated. Thus, we consider FP~\cite{liu2018finepruning} BNP~\cite{zheng2022bnp}, CLP~\cite{zheng2022clp}, ANP~\cite{wu2021anp}, AWN~\cite{chai2022awn}, RNP~\cite{li2023rnp}, MM-BD~\cite{wang2023mm}, NPD~\cite{zhu2024npd}, FT~\cite{liu2018finepruning}, FST~\cite{min2024fst}, FT-SAM~\cite{zhu2023ft-sam}, NAD~\cite{li2021nad}, NC~\cite{wang2019nc}, PBE~\cite{mu2023pbe}, i-BAU~\cite{zeng2022i-bau}, and SAU~\cite{wei2023sau}. We evaluate NC only using CIFAR-10 due to its computational complexity scaling with the number of classes. In addition, we do not include MESA~\cite{qiao2019mesa} and BAERASER~\cite{liu2022baeraser} because of their additional assumptions, such as knowledge of the target label and the approximate trigger position and size. Furthermore, we exclude BCU since it assumes access to an out-of-distribution dataset. While not a limitation, a fair evaluation of BCU would require benchmarking on multiple datasets, which is computationally prohibitive.

For approaches already implemented in BackdoorBench, we use the default configurations. We implement AWN, MM-BD, RNP, FST and PBE using the code provided by the authors, utilizing the training configurations reported in each paper. In cases where hyperparameter values are not explicitly reported, we use the values from the respective codebase. In the Supplementary Materials (Table~I), we summarise the hyperparameter values used for each approach.

\begin{figure*}[t]
    \centering
    \includegraphics[width=\linewidth]{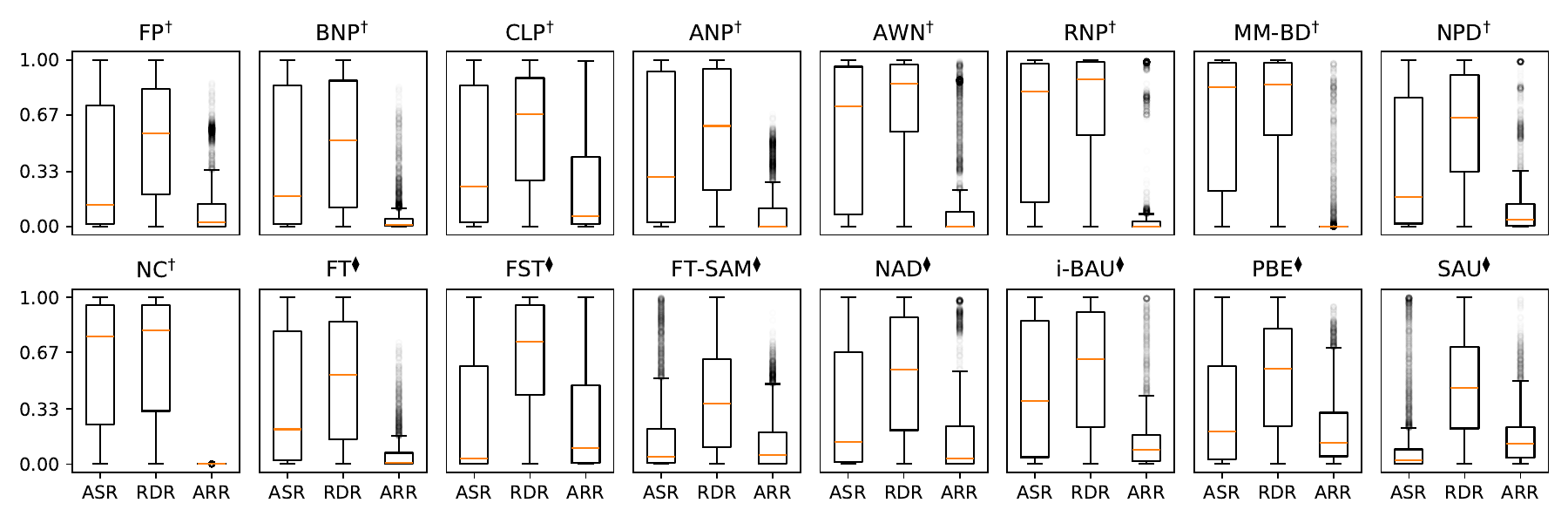}
    \caption{Box plots of the ASR, RDR, and ARR results for each approach across all considered settings. $\dagger$ = Pruning and $\blacklozenge$~=~Fine-tuning. Note: NC results are only for CIFAR-10.}
    \label{fig:result-summary}
\end{figure*}

\begin{figure}[t]
    \begin{subfigure}[b]{0.5\textwidth}
        \centering
        \includegraphics[width=0.98\textwidth]{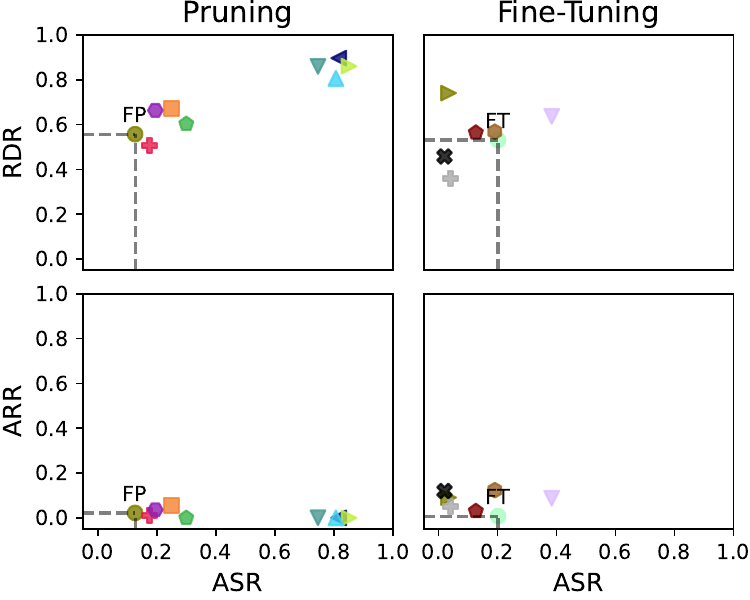}
    \end{subfigure}
    \begin{subfigure}[b]{0.5\textwidth}
        \centering
        \includegraphics[width=1.02\textwidth]{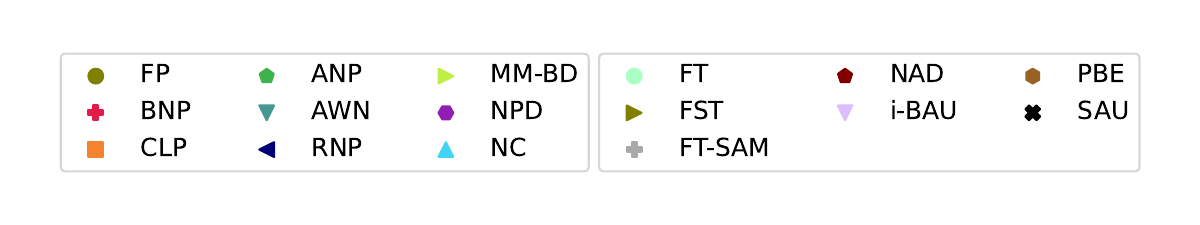}
    \end{subfigure}
    \caption{Scatter plot of the median RDR and ARR versus ASR for each approach across all considered settings. Note: NC results are only for CIFAR-10.}
    \label{fig:result-summary-2}
\end{figure}

\subsection{Other Settings}

We consider three datasets: CIFAR-10, German Traffic Sign Recognition Benchmark (GTSRB), and Tiny-ImageNet, containing 10, 43, and 200 classes, respectively. To evaluate the effect of data availability on the performance of each approach, we assess then under three data settings, based on the sample per class (SPC) value. Specifically, we evaluate each approach considering SPC values of 2, 10, and 100. For each data setting, we conduct 10 iterations, with each iteration utilizing a different random data partition. Moreover, we employ four model architectures: PreAct-ResNet18 (ResNet), VGG-19 with batch normalisation (VGG), EfficientNet-B3 (EfficientNet), and MobileNetV3-Large (MobileNet), using their default configurations as provided by \textit{BackdoorBench}.    

\subsection{Performance measures}

To evaluate the effectiveness of backdoor mitigation, we use three key performance measures, commonly used in the literature: clean accuracy (ACC), attack success rate (ASR), and recovery accuracy (RA). These metrics, though sometimes referred to by different names in various works, serve the same purposes:
\begin{itemize}
\item ACC represents the accuracy of the original classification task. It is measured as the accuracy of the model on the testing data without the backdoor trigger applied (i.e., the clean data).
\item ASR measures the effectiveness of the backdoor attack. It is calculated as the accuracy on testing data with the backdoor trigger applied (i.e., the backdoor data) and corresponding labels changed to the target label. Note that, testing data that originally belongs to the target class is omitted from this calculation. 
\item RA quantifies how effective a mitigation approach is at restoring the model's classification performance after backdoor mitigation. It measures the accuracy of testing data with the backdoor trigger applied but using the original (correct) labels.
\end{itemize}
In essence, ASR indicates whether the application of the trigger results in targeted misclassification, while RA shows whether the model can correctly classify backdoor samples after mitigation is applied, restoring them to their original labels. In our evaluations, we use normalised variants of ACC and RA, while ASR remains unchanged. These normalized metrics provide a clearer understanding of the model’s performance before and after the application of backdoor mitigation approaches. 

\subsubsection{Accuracy Reduction Ratio (ARR)}

To quantify the impact of mitigation on the accuracy of the original classification task, we take into account the accuracy values before and after mitigation, denoted as $\text{ACC}_{\text{pre}}$ and $\text{ACC}_{\text{post}}$, respectively. Therefore, we calculate the accuracy reduction ratio (ARR) as

\begin{equation} \label{arr-equation}
    \text{ARR} = \frac{\text{ACC}_{\text{pre}} - \text{ACC}_{\text{post}}}{\text{ACC}_{\text{pre}}}.
\end{equation}
Dividing the difference by the pre-mitigation accuracy accounts for variations in $\text{ACC}_{\text{pre}}$. For instance, most Tiny-ImageNet models exhibit lower $\text{ACC}_{\text{pre}}$ compared to their CIFAR-10 counterparts. Ideally, this ratio approaches $0$, indicating minimal reduction in accuracy due to mitigation. 

\subsubsection{Recovery Difference Ratio (RDR)}

To measure the effectiveness of a mitigation strategy in restoring the classification of backdoor samples to their original classes, we calculate the recovery difference ratio (RDR) as 

\begin{equation} \label{rdr-equation}
    \text{RDR} = \frac{\text{ACC}_{\text{pre}} - \text{RA}_{\text{post}}}{\text{ACC}_{\text{pre}}}.
\end{equation}
Similar to ARR, this ratio evaluates the difference between the post-mitigation RA and the pre-mitigation accuracy. In an optimal scenario, the post-mitigation RA, denoted by $\text{RA}_{\text{post}}$, ideally matches $\text{ACC}_{\text{pre}}$, while $\text{ACC}_{\text{post}}$ can often be impacted by the applied mitigation approach. As with ASR and ARR, a value of zero indicates an optimal outcome.

\section{Evaluation Results} \label{experimental-evaluation}

In this section, we present the results of our comprehensive evaluation of the considered backdoor attack mitigation approaches across various scenarios. We first discuss the overall performance of mitigation approaches across all experimental settings. Subsequently, we analyse the impact of data availability, backdoor attack type, model architecture, and dataset on the effectiveness of the examined mitigation approaches. 

\begin{figure*}[t]
    \centering
    \includegraphics[width=\linewidth]{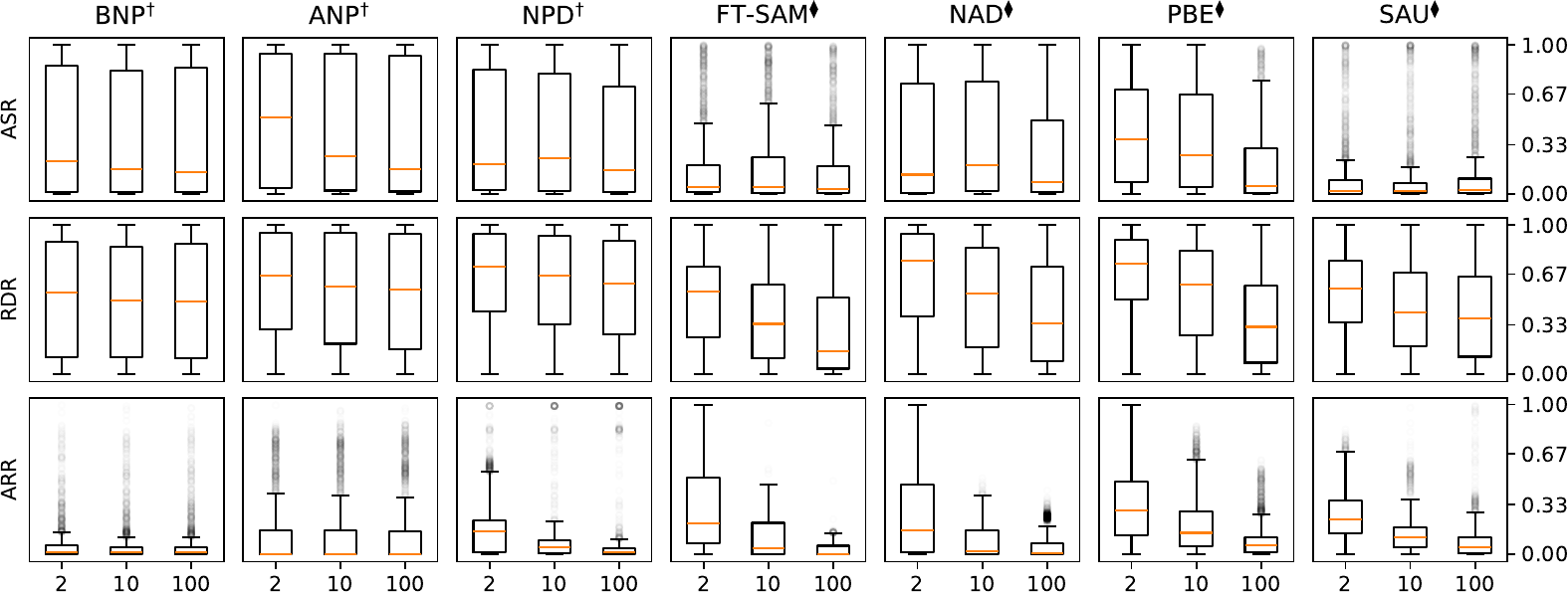}
    \caption{Box plots of the ASR, RDR, and ARR results for each approach and SPC values of 2, 10, and 100. $\dagger$ = Pruning and $\blacklozenge$~=~Fine-tuning. Note: CLP is omitted as it is data-free.}
    \label{fig:spc-summary}
\end{figure*}

\begin{figure}[t]
    \begin{subfigure}[b]{0.5\textwidth}
        \centering
        \includegraphics[width=0.98\textwidth]{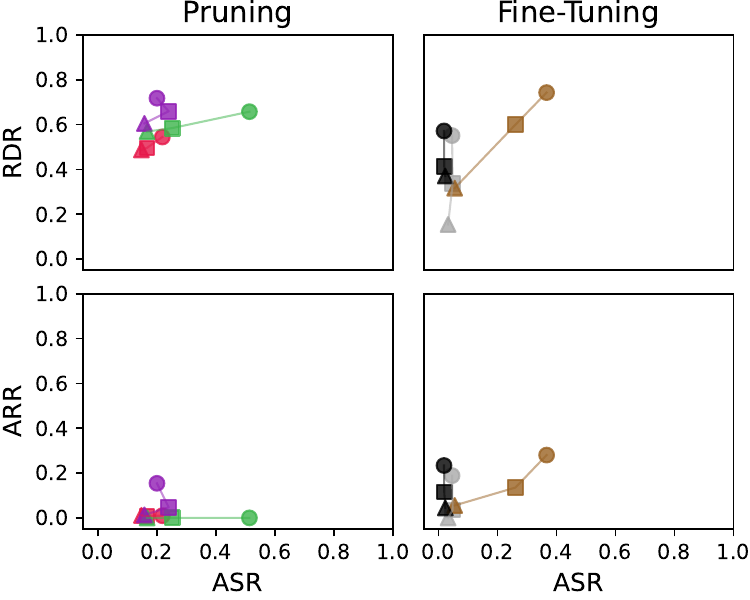}
    \end{subfigure}
    \begin{subfigure}[b]{0.5\textwidth}
        \centering
        \includegraphics[width=0.98\textwidth]{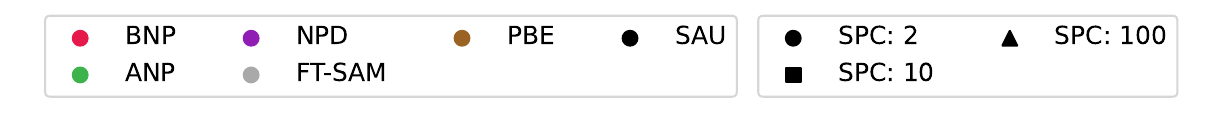}
    \end{subfigure}
    \caption{Scatter plots of the median RDR and ARR versus ASR for each approach and SPC values of 2, 10, and 100. Note: CLP is omitted as it is data-free.}
    \label{fig:spc-summary-2}
\end{figure}

\subsection{Overall Results} 

Figures~\ref{fig:result-summary} and \ref{fig:result-summary-2} present an overview of the results for each considered approach. In Fig.~\ref{fig:result-summary}, we use box plots to summarize the range of ASR, RDR, and ARR values, while, in Fig.~\ref{fig:result-summary-2}, we plot the median ARR and RDR of each approach against the respective median ASR. In Fig.~\ref{fig:result-summary}, the top row and NC are pruning approaches, while the second row, excluding NC, are fine-tuning approaches. Moreover, in Fig.~\ref{fig:result-summary-2}, we draw a rectangle using the median ARR, RDR, and ASR results of FP and FT. Approaches falling inside the bottom-left rectangle improve upon the performance of these baseline approaches. Note that the optimal performance in Fig.~\ref{fig:result-summary-2} corresponds to the bottom-left corner in all cases, and that FP and FT serve as the baseline pruning and fine-tuning approaches, respectively. Below, we discuss the pruning and fine-tuning approaches individually.

\subsubsection{Pruning Methods}

From Fig.~\ref{fig:result-summary-2}, it is evident that none of the evaluated approaches appear in the bottom left rectangle, defined by the median ARR, RDR, and ASR results of FP. Thus, none of the evaluated pruning approaches improve the median performance of FP. However, it is worth noting that ANP, BNP, CLP, and NPD perform comparably to FP across all three performance measures.

For metric-based pruning approaches (i.e., FP, BNP, and CLP), their overall effectiveness appears limited. Despite these approaches having low median ASR values, as shown in Fig.~\ref{fig:result-summary}, they all exhibit a heavy tail in the distribution of ASR results. In terms of ARR, both FP and BNP demonstrate good performance, with low medians and small variances. Although CLP has a median ARR comparable to FP and BNP, it displays significantly higher variance. In contrast, the ASR and RDR performance of NC is notably worse than FP, but it does not have the lowest median and variance for ARR among the evaluated approaches.

Masking-based pruning approaches (i.e., ANP, AWN, and RNP) also demonstrate limited overall effectiveness. ANP's distribution of ASR, RDR, and ARR values is similar to FP, but with increased variation in ASR and RDR. While RNP and AWN perform well in terms of ARR, their high median ASR and RDR offsets this benefit. When comparing ANP with AWN and RNP, we find that, although AWN and RNP have been designed to improve upon ANP, in our evaluations, ANP consistently outperforms both.

Additive pruning methods (i.e., MM-BD and NPD) also show limited effectiveness. Similar to RNP, MM-BD has high median ASR and RDR values. Although NPD appears to outperform MM-BD overall, it fails to improve upon FP in any performance measure.

\subsubsection{Fine-Tuning Methods}

In contrast to model pruning approaches, we find that FT-SAM and SAU outperform the baseline fine-tuning approach, FT. Both approaches fall within the RDR and ASR rectangle in Fig.~\ref{fig:result-summary-2} while also exhibiting a reduced ARR median with only a small ARR increase. On the other hand, FST, NAD, PBE, NC, and i-BAU do not surpass FT's performance, though NAD and, to some extent, PBE exhibit comparable performance to FT.

Among the conventional fine-tuning approaches (i.e., FST and FT-SAM), only FT-SAM outperforms FT, as indicated by its lower median ASR and RDR, as well as reduced variance, in Fig.~\ref{fig:result-summary}. However, this improvement comes at the expense of a higher median ARR and increased variance. While FST achieves slightly reduced ASR and ARR relative to FT, underperforms in terms of RDR. Additionally, FST exhibits higher variance in ASR and ARR compared to FT-SAM. NAD demonstrates overall performance comparable to FT but with increased variance in ARR.


Adversarial training approaches (i.e., PBE, i-BAU, and SAU) exhibit varied performance. Compared to FT, i-BAU shows lower performance across all three measures. PBE achieves similar ASR and RDR compared to FT, though it exhibits a higher median ARR and a longer tail. In contrast, SAU demonstrates significant improvement over FT, with a lower median ASR and reduced variance. However, similar to FT-SAM, this improvement comes with a trade-off in ARR performance. 

\subsubsection{Summary}

With the exception of SAU and FT-SAM, all evaluated approaches exhibit variable performance across the full range of tested settings. Most approaches demonstrate considerable variability in ASR and RDR, highlighting the need for caution when applying these approaches in real-world scenarios. Nonetheless, both SAU and FT-SAM achieve low median ASR with reduced variance, making them more suitable for practical applications, despite a trade-off in ARR performance and continued poor RDR performance. Lastly, the overall improvement compared to FP and FT, introduced in 2018, is less substantial than claimed in most works. Future research can benefit from focusing on enhancing RDR performance, as it remains a significant challenge for many existing approaches.

In the following subsections, we discuss BNP, CLP, ANP, and NPD as they are the best-performing pruning approaches. Similarly, we discuss FT-SAM, NAD, PBE, and SAU as the top-performing fine-tuning approaches. We provide the complete set of results in the Supplementary Materials.

\begin{figure*}[t]
    \centering
    \includegraphics[width=\linewidth]{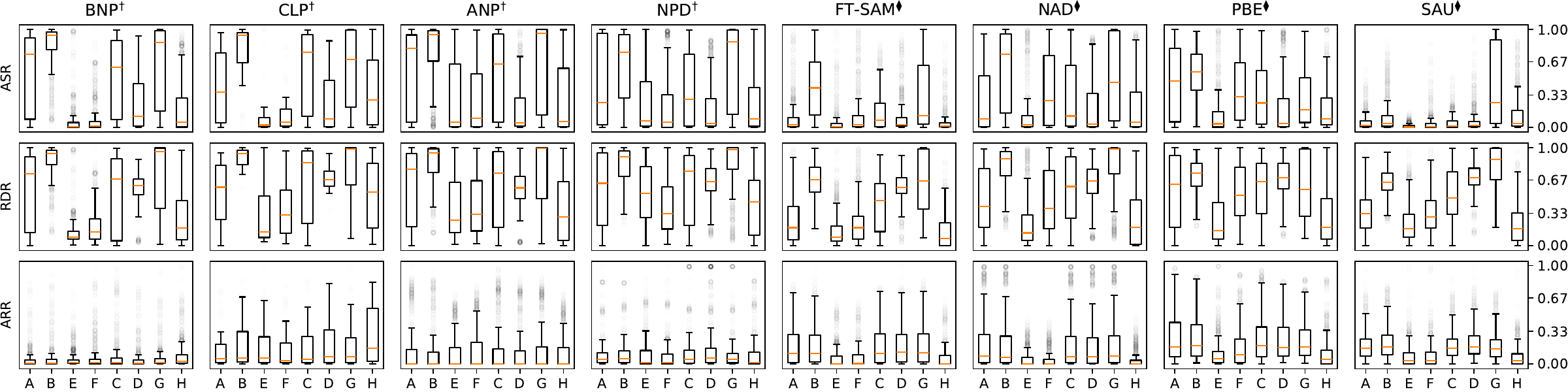}
    \caption{Box plots of the ASR, RDR, and ARR results for the selected approaches and different attack types. $\dagger$ = Pruning and $\blacklozenge$~=~Fine-tuning. A = BadNet, B = Blended, C = LF, D = Signal, E = BPP, F = IAB, G = SSBA and H = WaNet.}
    \label{fig:attack-summary}
\end{figure*}

\begin{figure}[t]
    \centering
    \includegraphics[width=\linewidth]{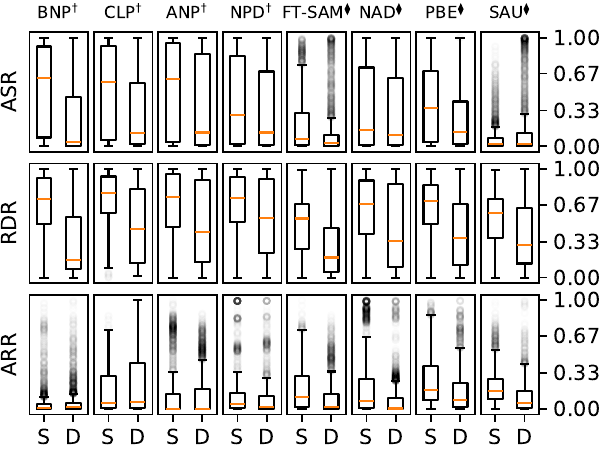}
    \caption{Box plots of the ASR, RDR, and ARR results for the selected approaches and both static and dynamic attacks. $\dagger$ = Pruning and $\blacklozenge$~=~Fine-tuning. S = Static and D = Dynamic.}
    \label{fig:attack-type-summary}
\end{figure}

\subsection{Data Availability}

To assess the effect of data availability, we evaluate each approach using 2, 10, and 100 samples per class (SPC). Figs.~\ref{fig:spc-summary} and \ref{fig:spc-summary-2} show the SPC results of the best-performing approaches, as identified in the previous section. In Fig.~\ref{fig:spc-summary}, we summarize the range of ASR, RDR, and ARR values using box plots, while in Fig.~\ref{fig:spc-summary-2}, we plot the median ARR and RDR of each approach against the median ASR. We provide the complete set of results in the Supplementary Materials (Fig.~1). Overall, we observe that data availability impacts ASR, RDR, and ARR, as indicated by the square and triangle points in Fig.~\ref{fig:spc-summary-2} diverging from the bottom-left corner. However, pruning-based approaches tend to perform more consistently when SPC is reduced.  

\subsubsection{Model-Pruning}

Data availability appears to have a minimal impact on the performance of most pruning methods. In particular, BNP and ANP demonstrate consistent ARR and RDR performance. However, the median ASR for ANP increases as SPC is reduced. In contrast, NPD is more significantly affected by a reduction in SPC. While there are only minor differences in ASR and RDR, a noticeable increase in ARR median occurs as SPC decreases.

\subsubsection{Fine-Tuning} 

Unlike the evaluated pruning approaches, SPC has a greater impact on the performance of fine-tuning approaches. Specifically, median ARR and RDR increase significantly for FT-SAM, PBE, and NAD when SPC is reduced. Although a similar trend is observed for SAU, the impact of SPC is less pronounced.

\subsection{Backdoor Attack} \label{results:attack}

In Fig.~\ref{fig:attack-summary}, we present the results for the select set of approaches across the eight considered backdoor attacks. We provide the full set of results in the Supplementary Materials (Fig.~2 and 3). The performance of most approaches varies significantly across the range of tested attacks, particularly in terms of ASR and RDR. For example, although BNP performs well against the BPP attack, its performance against the Blended attack is the worst among all approaches. In contrast, FT-SAM and SAU demonstrate more consistent performance across the attacks. Notable exceptions are SSBA and WaNet for SAU, and Blended and SSBA for FT-SAM. However, it is important to note that FT-SAM and SAU's RDR performance fluctuates similarly to those of other approaches.

In Fig.~\ref{fig:attack-type-summary}, we group the results from Fig.~\ref{fig:attack-summary} by attack type, as categorised in Table~\ref{tab:attack-details}. We observe that dynamic backdoor attacks are better mitigated by most approaches, as indicated by the lower median ASR and RDR values. Interestingly, several recent approaches, including NPD, PBE, and SAU, specifically target this type of attack.

Among the tested attacks, Blended and SSBA are the most difficult to defend against, as evidenced by higher median ASR median values and greater variance across most approaches. Additionally, the ASR and RDR results for the BadNets attack vary significantly for most approaches, with exception of FT-SAM and SAU. This is surprising given that BadNets is considered the foundational attack. 

\subsection{Model Architecture}

\begin{figure*}[t]
    \centering
    \includegraphics[width=\linewidth]{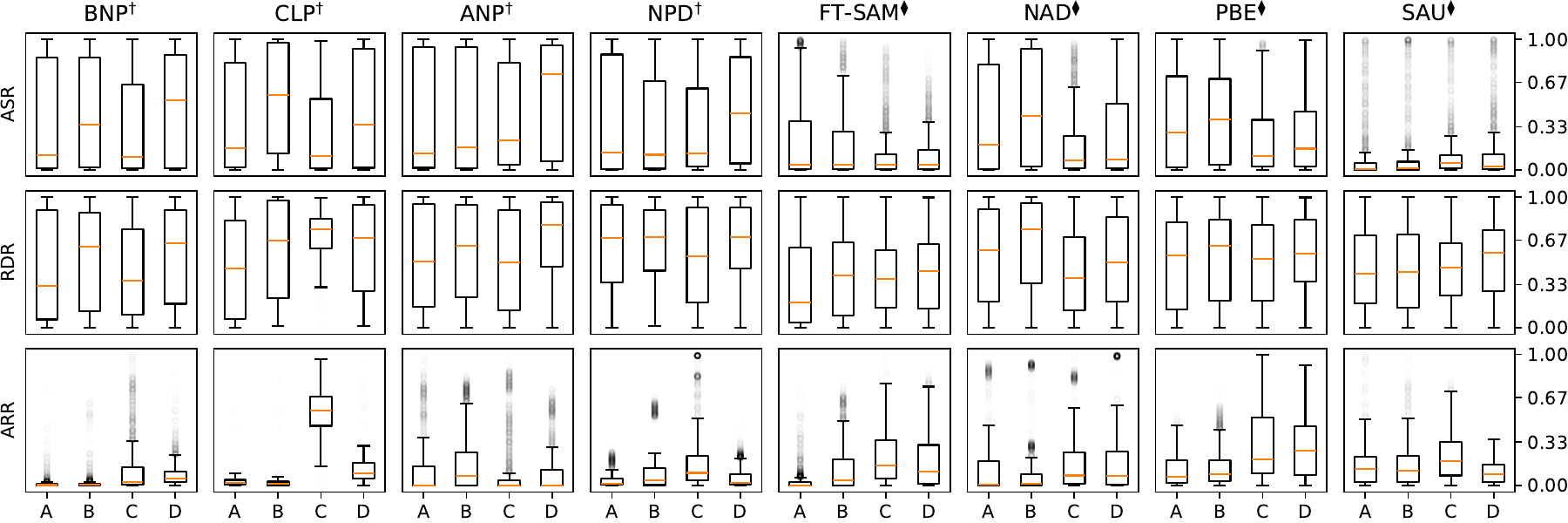}
    \caption{Box plots of the ASR, RDR and, ARR results for the selected approaches and different model architectures. $\dagger$ = Pruning and $\blacklozenge$~=~Fine-tuning. A = VGG, B = ResNet, C = EfficientNet and D = MobileNet.}
    \label{fig:model-summary}
\end{figure*}

Fig.~\ref{fig:model-summary} shows the results for each selected approach across the four tested model architectures. We provided the results for all approaches in the Supplementary Materials (Fig.~4). Except for SAU, most approaches exhibit inconsistent performance across considered architecture types, as indicated by fluctuations in ASR or ARR values. Notably, SAU demonstrates the most consistent performance across all architectures.

For CLP, median ARR increases significantly when using the EfficientNet architecture, with substantial fluctuations in median ASR and RDR values. Similarly, although BNP's ARR performance remains mostly stable, its median RDR and ASR show considerable variation. For ANP and NPD, median ASR noticeably increases when the MobileNet architecture is used. 

While FT-SAM maintains relatively consistent median ARR and ASR values across considered architectures, the tails of the distributions of these measures expand in certain cases. Moreover, there appears to be an inverse relationship between the tails of FT-SAM's ARR and ASR distributions. That is, a reduction in the size of the ARR distribution tail is often accompanied by an increase in the length and weight of the ASR distribution tail, and vice versa. In contrast, SAU exhibits only minor differences in all three performance measures across considered architectures.

\subsection{Dataset} 

\begin{figure*}[t]
    \centering
    \includegraphics[width=\linewidth]{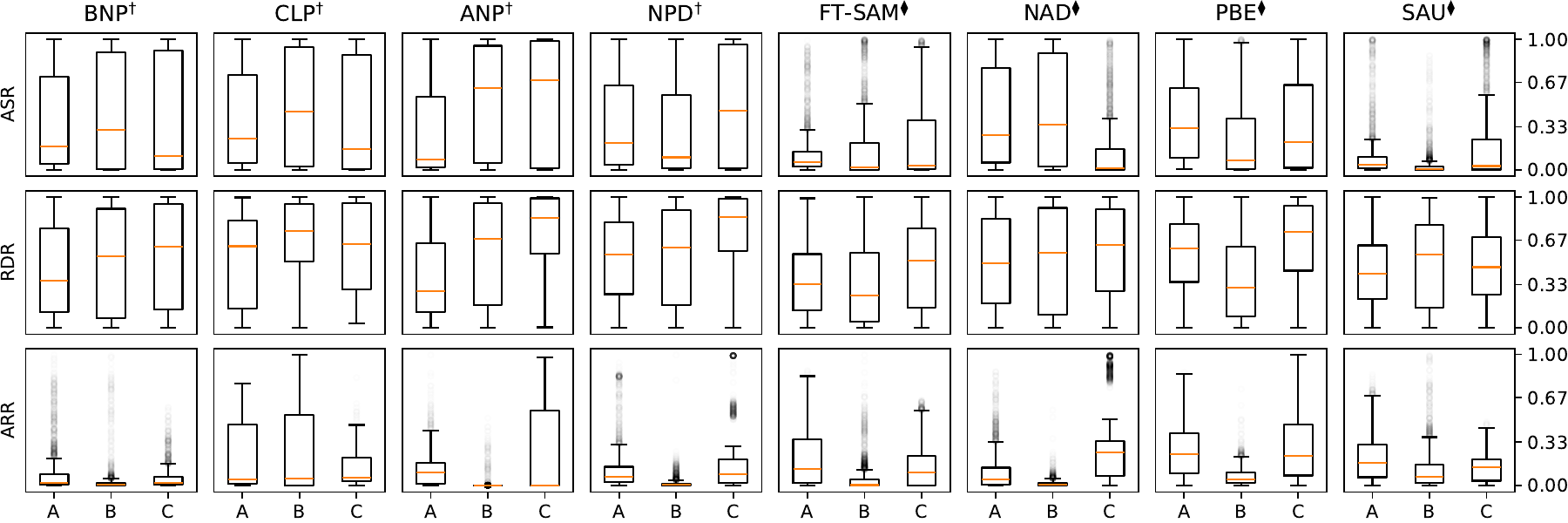}
    \caption{Box plots of the ASR, RDR, and ARR results for the selected approaches and different datasets. $\dagger$ = Pruning and $\blacklozenge$~=~Fine-tuning. A = CIFAR-10, B = GTSRB and C = Tiny-ImageNet.}
    \label{fig:dataset-summary}
\end{figure*}

Fig.~\ref{fig:dataset-summary} shows the results for each selected approach across the three considered datasets. We provided the complete results in the Supplementary Materials (Fig.~5). Similar to the performance variability observed with different model architectures, there is noticeable variability across datasets, often more pronounced. 

Except for BNP, each approach exhibits variations in median or distribution tail of ARR across datasets. In particular, ANP shows significant variation in tail weight for the GTSRB and Tiny-ImageNet tail, a trend also observed in other approaches. For RDR, increasing the complexity of the classification task generally leads to worse performance, with CLP and SAU being exceptions. This is evident from the increase in median RDR from CIFAR-10 to GTSRB and from GTSRB to Tiny-ImageNet. A similar trend is present for ASR, where ANP, FT-SAM, NPD, and SAU perform worse as task complexity increases.

\subsection{Poisoning Ratio} 

\begin{figure*}[t]
    \centering
    \includegraphics[width=\linewidth]{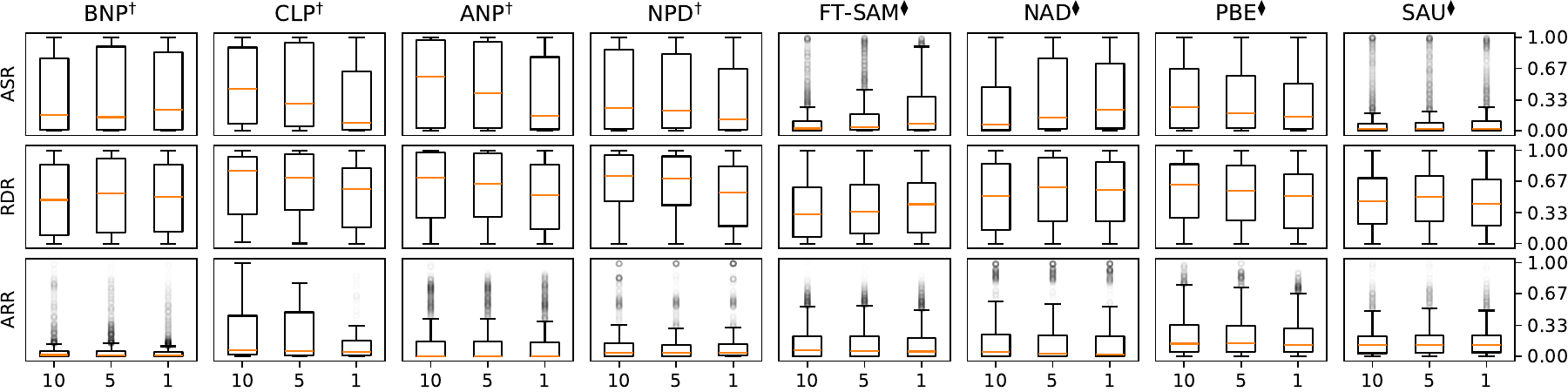}
    \caption{Box plots of the ASR, RDR, and ARR results for the selected approaches and different poisoning ratios (\%). $\dagger$ = Pruning and $\blacklozenge$~=~Fine-tuning.}
    \label{fig:p-ratio-summary}
\end{figure*}

Fig.~\ref{fig:p-ratio-summary} shows the results for each selected approach across the three tested poisoning ratios. We present the full set of results in the Supplementary Materials (Fig.~6). Unlike the findings in~\cite{wu2022backdoorbench}, we observe that the poisoning ratio does not significantly affect the performance of most approaches. While some differences are noticeable, these variations are less pronounced compared to other variables discussed thus far. However, it is worth noting that the ASR results of FT-SAM are significantly impacted when the poisoning ratio is reduced.

\section{Discussion}

In this section, we discuss the major findings of our survey, connecting the literature reviewed in sections~\ref{model-pruning} and \ref{fine-tuning} with the evaluation results in section~\ref{experimental-evaluation}.

\begin{table}[]
    \centering
    \caption{ Summary of the characteristics of the discussed approaches. \blackCircle: Multi-scenario evaluation, \halfBlackCircle: Single scenario evaluation and \whiteCircle: No evaluation.}
    \scalebox{0.9}{
    \begin{tabular}{cc|ccc|cc}
        \toprule
        \multirow{2}{*}{Ref} & \multirow{2}{*}{Name} & \multicolumn{3}{c|}{Trigger Modeling} & \multicolumn{2}{c}{Hyperparameters} \\
        & & Use & Type & Norm & \# $\lambda$ & Analysis \\
        \toprule
        \cite{liu2018finepruning} & FP & \xmark & - & - & 0 & - \\
        \cite{zheng2022bnp} & BNP & \xmark & - & - & 1 & \halfBlackCircle \\
        \cite{zheng2022clp} & CLP & \xmark & - & - & 1 & \blackCircle \\
        \cite{wu2021anp} & ANP & \xmark & - & - & 1 & \blackCircle \\
        \cite{chai2022awn} & AWN & \cmark & Global & L2 & 3 & \halfBlackCircle \\
        \cite{li2023rnp} & RNP & \xmark & - & - & 0 & - \\
        \cite{wang2023mm} & MM-BD & \xmark & - & - & 1 & \halfBlackCircle \\
        \cite{zhu2024npd} & NPD & \cmark & Sample Specific & L2 & 3 & \whiteCircle \\
        \cite{min2024fst} & FST & \xmark & - & - & 1 & \halfBlackCircle  \\
        \cite{zhu2023ft-sam} & FT-SAM & \xmark & - & - & 0 & - \\
        \cite{li2021nad} & NAD & \xmark & - & - & 1 & \blackCircle \\
        \cite{pang2023bcu} & BCU & \xmark & - & - & 0 & - \\
        \cite{qiao2019mesa} & MESA & \cmark & Global & L1 & 1 & \whiteCircle \\
        \cite{liu2022baeraser} & BAERASER & \cmark & Global & L1 & 2 & \blackCircle \\
        \cite{wang2019nc} & NC & \cmark & Global & L1 & 1 & \whiteCircle \\
        \cite{mu2023pbe} & PBE & \cmark & Sample Specific & L2 & 0 & - \\
        \cite{zeng2022i-bau} & i-BAU & \cmark & Global & L2 & 0 & - \\
        \cite{wei2023sau} & SAU & \cmark & Sample Specific & L2 & 3 & \whiteCircle \\       
    \hline
    \end{tabular}}
    \label{tab:additional-analysis}
\end{table}

\subsection{Sensitivity}

Most of the evaluated approaches exhibit highly variable performance when tested across a broader set of scenarios compared to those originally considered in the respective papers. Although it is unrealistic to expect any particular approach to have invariant performance, the variance observed in most approaches is considerable. Specifically, ASR and RDR show significant variability in most cases. Among the evaluated variables, data availability, attack type, and dataset have the largest impact on performance. 

For data availability, fine-tuning approaches are notably affected when SPC is reduced. Particularly, the performance of the original classification task, as quantified by ARR, deteriorates with reduced SPC. This is a significant observation, as a substantial decrease in ARR can render an approach impractical for real-world applications, irrespective of the removal of the backdoor. The inherent complexity of the optimization problems proposed by each fine-tuning approach often results in low bias but high variance, which increases the likelihood of overfitting to $\mathcal{D}_{m}$ when SPC is reduced. Although many approaches attempt to alleviate this issue by constraining the weight or input perturbation space or by defining a multi-objective optimization problem where one objective is to preserve the performance on $\mathcal{D}_{m}$, these approaches do not effectively constraint $\theta$ to account for overfitting.

Although FST constrains the solution space of $\theta$, its minimisation of the inner product encourages differences among parameters. In contrast, while FT-SAM restricts the perturbation applied to $\theta$, the outer minimisation imposes no direct constraint to $\theta$. Given that a practical backdoor mitigation procedure typically operates under the assumption of limited access to clean data, it is crucial for future fine-tuning methods to develop effective strategies that constrain $\theta$ to mitigate the risk of overfitting. By achieving this, we posit that the mitigation task will become more tractable and better positioned to address the bias-variance trade-off while maintaining the performance of the original classification task.

Regarding attack type, dataset, and model architecture, most approaches show variable performance in removing and restoring the backdoor task, as measured by ASR and RDR. Unlike ARR, both ASR and RDR are inaccessible to the defender after mitigation, making large uncertainty in these measures a significant risk for practical use. 

The sensitivity of the performance of most approaches to the values of their hyperparameters is a major factor contributing to this variability. While some approaches evaluate this sensitivity in different settings (cf. Table~\ref{tab:additional-analysis}), the majority of evaluations are often limited in extent due to practical constraints. This underscores that although hyperparameters are often unavoidable, their inclusion requires careful consideration. Unlike traditional deep learning applications, using validation datasets to find optimal hyperparameter values is impractical for two main reasons. First, as mentioned earlier, ASR and RDR are not observable by the defender in practice, making hyperparameter tuning effective only for minimising ARR. Second, the use of a validation dataset further limits the available mitigation data, which can exacerbate the impact of data availability on overall performance.

Another factor contributing to performance variance across attack types, datasets, and model architectures is the reliance on limited observational evidence. Specifically, most backdoor mitigation strategies are built upon analyzing backdoored models and characterizing their behaviour. For instance, CLP investigates the correlation between UCLC and TAC, employing UCLC to prune filters with outlier values. However, our benchmarking results indicate that none of these observations can be deemed universal characteristics associated with backdoor attacks given the varied performance of most proposals. Rather, these observations are likely indicative of backdoor attacks within the specific settings tested by each approach. While we recognise the practical implications of examining the range of settings evaluated in this work, our findings underscore the importance of considering variations in attack type, dataset, and model architecture.

Future mitigation approaches ought to carefully account for the impact of limited data availability on performance. Specifically, a more deliberate consideration of the bias-variance trade-off is essential when designing optimisation-based approaches. In addition, the inclusion of hyperparameters demands careful thought, as their values can significantly affect practical performance. Since hyperparameter optimization is often challenging or impractical in real-world scenarios, sensitivity to hyperparameter values becomes even more critical in such scenarios. Finally, future investigations into the underlying mechanisms that drive backdoor attacks (assuming such mechanisms exist) need to carefully account for variations in attack type, dataset, and model architecture.

\subsection{Recovery Accuracy} \label{discussion:ra}

Among the surveyed works, we find that only the proposers of SAU evaluate the ability of their mitigation approach to restore the classification of the backdoor task, quantified as RDR in this paper. While ASR measures how effectively the backdoor induces the misclassification of samples containing the trigger to the target class, BackdoorBench highlights that this measure alone does not determine the overall effectiveness of a mitigation approach. Specifically, minimizing ASR without a corresponding increase in RA, the performance measure used to calculate RDR, is not indicative of optimal performance. If ASR is minimised but RA remains low, the model is still unable to accurately classify samples containing the trigger. Although this may not align with the adversary's original objective, it still has significant implications for models deployed in real-world settings.

Across the tested settings, RDR varies significantly. Even though FT-SAM and SAU are state-of-the-art approaches, their RDR performance still exhibit notable variability. While RDR is dependent on ASR and ARR, as $\text{ASR} + \text{RA} \leq 1$ and $\text{RA} \lessapprox \text{ACC}$, the median and variance of RDR often exceed the values expected from the relationship between these performance measures (see Figs.~\ref{fig:result-summary} and \ref{fig:result-summary-2}). 

Since samples containing the trigger are inaccessible to the defender, restoring the classification of the backdoor task presents a major challenge. Despite efforts by the proposers of approaches NPD, AWN, NC, MESA, BAERASER, PBE, i-BAU, and SAU to model the trigger distribution, a substantial improvement in RDR has not been observed. Therefore, moving forward, more focused exploration of alternative methodologies targeting RDR is necessary. 

\subsection{Trigger Modeling}

Modelling the trigger distribution is a widely adopted technique, used in nearly half of the surveyed works (cf. Table~\ref{tab:additional-analysis}). Approaches that utilize this technique rely on the insights derived from the trigger model to mitigate corresponding backdoor attacks. A key feature shared among these approaches is the use of a constrained optimisation to determine $\delta$, though the specific implementation details differ across approaches. This constrained optimisation requires selecting a norm and an upperbound ($\epsilon$) for the norm of $\delta$. This inherently involves certain assumptions about $\rho$, the actual trigger employed by the adversary.

Among the discussed approaches, NC, MESA, and BAERASER constrain the $\ell_1$ norm of $\delta$, while NPD, AWN, PBE, i-BAU, and SAU utilize the $\ell_2$ norm. In the evaluated attacks, triggers employed by the BadNets and, arguably, IAB are sparse in nature, whereas other attacks apply smoother, less perceptible triggers. Interestingly, our results suggest that using an $\ell_1$ norm constraint in trigger modelling does not significantly enhance mitigation performance against attacks with sparse triggers compared to using an $\ell_2$ norm constraint. Furthermore, we observe that constraining the $\ell_2$ norm does not guarantee successful mitigation against attacks with smoother triggers, as illustrated by the results for the Blended and SSBA attacks in section~\ref{results:attack}.

One of the main challenges with using the $\ell_2$ norm is the natural occurrence of adversarial examples within the input space. To tackle this, AWN and i-BAU model $\delta$ as a global input perturbation. However, the results for AWN and i-BAU indicate that this additional constraint leads to suboptimal performance. In contrast, NPD, PBE, and SAU adopt a sample-specific approach to modelling the trigger distribution.
NPD assigns a unique $\delta$ for each $x$ based on its second-largest logit, ensuring that the perturbation $\delta$ causes targeted misclassification. However, our findings in section~\ref{results:attack} reveal that this selection method yields variable performance. Conversely, PBE employs an existing untargeted adversarial example generation method, demonstrating that untargeted adversarial examples generated using the PGD attack tend to exhibit biased classification towards the adversary's target class. However, our findings suggest that the error rate of the PGD attack (i.e., the proportion of adversarial examples not classified as the target class) impacts PBE's performance, particularly in scenarios with limited data availability. 

To account for the presence of adversarial examples in sample-specific trigger modelling with $\ell_2$ norm constraint, SAU filters candidate triggers using the original model parameters. Specifically, SAU identifies sample-specific perturbations $\delta$ that induce consistent misclassification given the original model parameters $\theta$ and the modified ones $\bar{\theta}$. Our results show that this strategy more effectively distinguishes between adversarial examples and candidate triggers, as evidenced by robust ASR performance across the tested attacks. However, it is important to note that this approach alone is insufficient for restoring the correct classification of backdoor samples, as discussed earlier (see section~\ref{discussion:ra}).

\section{Conclusion}

We critically evaluated various state-of-the-art backdoor attack mitigation strategies within the context of image recognition. Our analysis, spanning a broad spectrum of attacks, datasets, model architectures, data availabilities, and poisoning ratios uncovers several key insights into the effectiveness and limitations of current approaches:
\begin{itemize}
    \item While many approaches demonstrate strong performance in specific settings, most exhibit significant variability, particularly when faced with diverse attack types and constrained data availability. Pruning-based approaches such as BNP, ANP, and CLP offer a degree of robustness but struggle with performance consistency. Fine-tuning approaches, notably FT-SAM and SAU, show promise by outperforming their respective baselines, though they come with trade-offs, especially in terms of accuracy reduction and recovery performance.
    \item The widespread reliance on hyperparameters and constrained optimization techniques introduces significant challenges, particularly in real-world deployment. While hyperparameters are crucial to the success of many approaches, they must be carefully tuned to avoid overfitting and ensure robustness across varied attack scenarios. Balancing the bias-variance trade-off in optimization-based approaches is critical for future improvements.
    \item Trigger modeling remains a key technique in mitigating backdoor attacks. However, our results suggest that common assumptions about the trigger distribution do not universally hold across all attack scenarios. This highlights the need for more adaptive approaches that can account for variations in attack types and the properties of input data.
    \item A major challenge across all evaluated approaches is the restoration of backdoor-affected classifications, as quantified by RA and RDR. Most approaches, despite reducing attack success rates, fail to fully recover the correct classification of backdoor samples, underscoring the need for future research to focus on improving RDR.
\end{itemize}
In summary, while considerable progress has been made in developing backdoor mitigation strategies, our findings highlight the need for more adaptive, robust, and generalizable solutions. Future research can focus on addressing trade-offs between accuracy, recovery, and computational efficiency, while also exploring new approaches to mitigate backdoor attacks in diverse real-world scenarios.

\bibliographystyle{IEEEtran}
\bibliography{ref}
\vspace{-35pt}
\begin{IEEEbiography}[{\includegraphics[width=1in,height=1.25in,clip,keepaspectratio]{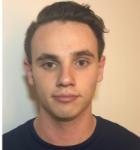}}]{Kealan Dunnett}
is a researcher at the School of Computer Science, Queensland University of Technology, Brisbane, Australia and the Commonwealth Scientific and Industrial Research Organization (CSIRO), Pullenvale, QLD, Australia. Kealan has several publications in highly ranked conferences/workshops including IEEE TrustCom, IEEE ICCCN, IEEE DSN, IEEE Internet of Things Magazine.
\end{IEEEbiography}
\vspace{-35pt}
\begin{IEEEbiography}[{\includegraphics[width=1in,height=1.25in,clip,keepaspectratio]{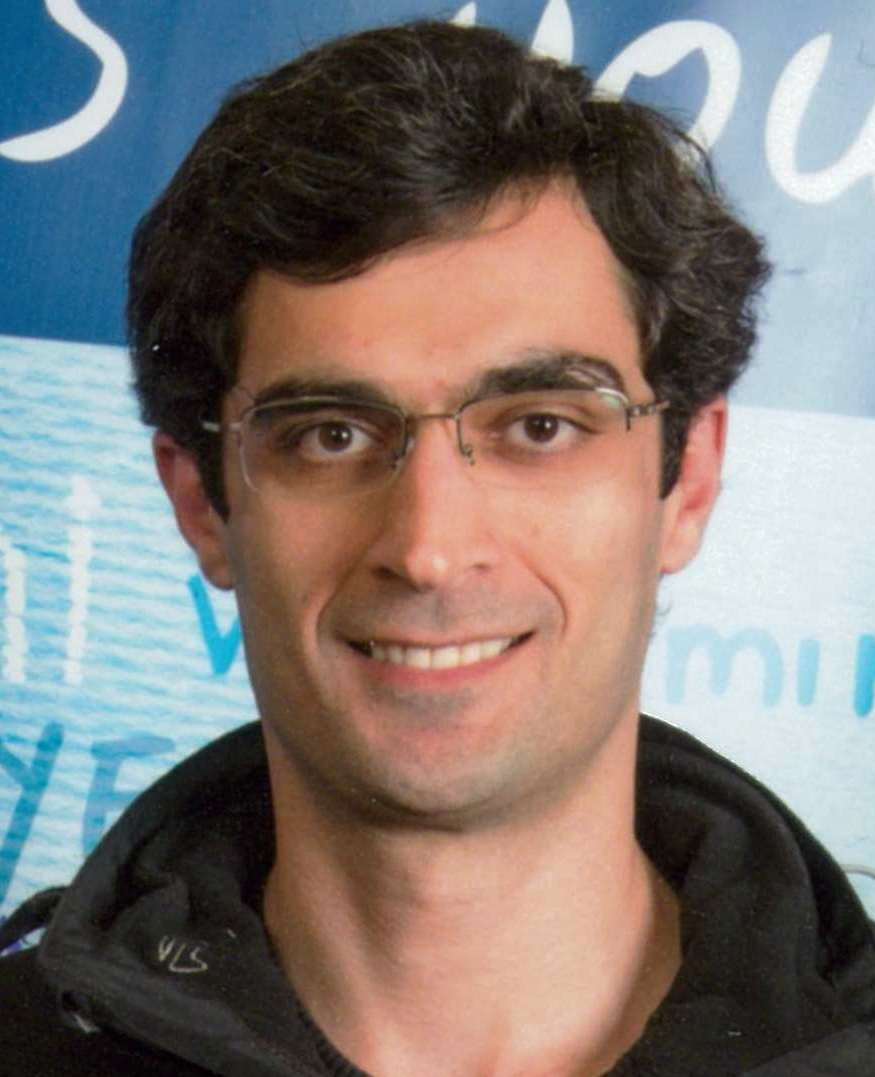}}]{Reza Arablouei}
received the Ph.D. degree in telecommunications engineering from the Institute for Telecommunications Research, University of South Australia, Mawson Lakes, SA, Australia, in 2013. He was a Research Fellow with the University of South Australia from 2013 to 2015. Since 2015, he has been with the Commonwealth Scientific and Industrial Research Organization (CSIRO), Pullenvale, QLD, Australia with the current position of Senior Research Scientist. His current research interests are around signal processing and machine learning on embedded systems.
\end{IEEEbiography}
\vspace{-35pt}
\begin{IEEEbiography}[{\includegraphics[width=1in,height=1.25in,clip,keepaspectratio]{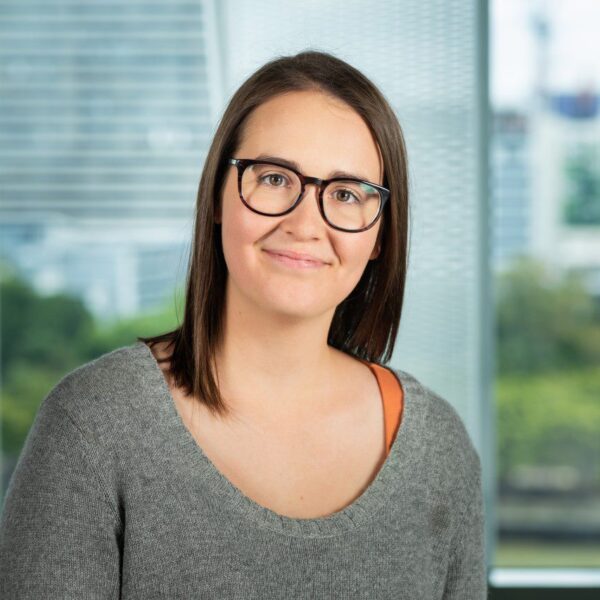}}]{Dimity Miller}
Dimity Miller received the Ph.D. degree in uncertainty in deep learning for robotic vision from QUT in 2021. She is a Chief Investigator with the Centre for Robotics, QUT, where she is also a Lecturer with the School of Electrical Engineering and Robotics. In 2021, she was with the ARC Centre of Excellence for Robotic Vision. Prior to joining the Centre for Robotics, QUT, she was a Post-Doctoral Research Fellow jointly across the CSIRO Machine Learning and Artificial Intelligence Future Science Platform, the CSIRO Robotics and Autonomous Systems Group, and the QUT Trusted Networks Laboratory. Her research expertise is in reliable robotic vision operating at the intersection of deep learning, computer vision, and robotics. Her thesis was recognized by an Executive Dean’s Commendation for Outstanding Doctoral Thesis Award in 2022.
\end{IEEEbiography}
\vspace{-35pt}
\begin{IEEEbiography}[{\includegraphics[width=1in,height=1.25in,clip,keepaspectratio]{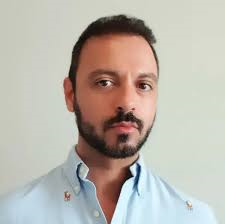}}]{Volkan Dedeoglu}
is a research scientist in the Distributed Sensing Systems Group of CSIRO Data61. His current research focuses on data trust and blockchain-based IoT security and privacy. Volkan also holds Adjunct Lecturer positions at UNSW Sydney and QUT. Before joining to Data61, Volkan worked as a postdoctoral researcher at Texas A\&M University on physical layer security for communications. He received his PhD in Telecommunications Engineering from UniSA (Australia, 2013), MSc in Electrical and Computer Engineering from Koc University (Turkey, 2008), and BSc in Electrical and Electronics Engineering from Bogazici University (Turkey, 2006).
\end{IEEEbiography}
\vspace{-35pt}
\begin{IEEEbiography}[{\includegraphics[width=1in,height=1.25in,clip,keepaspectratio]{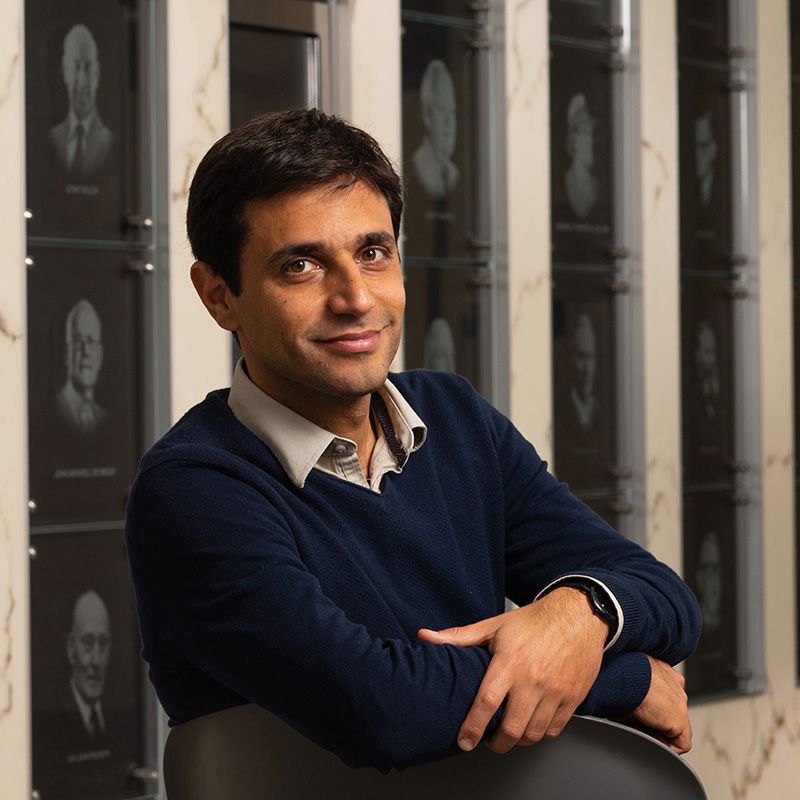}}]{Raja Jurdak} (Senior Member, IEEE) received the M.S. and Ph.D. degrees from the University of California at Irvine. He is a Professor of distributed systems and the Chair of applied data sciences with Queensland University of Technology, and the Director of the Trusted Networks Laboratory. He is also an Adjunct Professor with the University of New South Wales and a Visiting Researcher with Data61, CSIRO. Previously, he established and led the Distributed Sensing Systems Group, Data61, CSIRO. He also spent time as a Visiting Academic with MIT and Oxford University in 2011 and 2017, respectively. He has published over 250 peer reviewed publications, including three authored books on IoT, blockchain, and cyberphysical systems. His publications have attracted over 14,700 citations, with an H-index of 54. His research interests include trust, mobility, and energy efficiency in networks. He serves on the organizing and technical program committees of top international conferences, including Percom, ICBC, IPSN, WoWMoM, and ICDCS. He serves on the editorial boards of Ad Hoc Networks and Scientific Reports (Nature). He is an IEEE Computer Society Distinguished Visitor.
\end{IEEEbiography}

\end{document}